\documentclass[fleqn,usenatbib]{mnras}

\usepackage{newtxtext,newtxmath}

\usepackage[T1]{fontenc}

\DeclareRobustCommand{\VAN}[3]{#2}
\let\VANthebibliography\thebibliography
\def\thebibliography{\DeclareRobustCommand{\VAN}[3]{##3}\VANthebibliography}

\usepackage{graphicx}	
\usepackage{amsmath}	
\usepackage{subcaption}
\usepackage{multirow}

\newcommand{\Msun}{\ensuremath{\,{\rm M}_\odot}}                  
\newcommand{\kms}{\,km\,s$^{-1}$}                                 

\graphicspath{ {images/} }
\newcommand{\Msunnom}{\hbox{$\mathcal{M}^{\rm N}_\odot$}}
\newcommand{\Rsunnom}{\hbox{$\mathcal{R}^{\rm N}_\odot$}}
\newcommand{\Lsunnom}{\hbox{$\mathcal{L}^{\rm N}_\odot$}}

\title[VV Ori]{The enigmatic multiple star VV Ori }

\author[Budding et al.]
    {Edwin Budding$^{1,2}$,
    John Southworth$^{3}$\thanks{Corresponding author: taylorsouthworth@gmail.com},
    Kre\v{s}imir Pavlovski$^{4}$,
    Michael D.\ Rhodes$^{5}$,
    Wu Zihao$^{6}$, 
     \newauthor 
    Tom Love$^{7}$,   
    Mark G.\ Blackford$^{8}$,
    Timothy S.\ Banks$^{9,10}$, \&
    Murray Alexander$^{11}$
\vspace{2mm} \\
$^{1}$Carter Observatory, {40 Salamanca Road, Kelburn, Wellington 6012, New Zealand}\\
$^{2}$School of Chemical \& Physical Sciences, Victoria University of Wellington, PO Box 600, Wellington 6140, New Zealand\\
$^{3}$Astrophysics Group, Keele University, Staffordshire, ST5 5BG, UK\\
$^{4}$Department of Physics, Faculty of Science, University of Zagreb, Bijeni\v{c}ka cesta 32, 10000 Zagreb, Croatia\\
$^{5}${Brigham Young University, Provo, Utah 84602,} USA\\
$^{6}$Dept.\ Statistics \& Data Science, National University of Singapore, 6 Science Drive 2, Singapore 117546\\
$^{7}$Variable Stars South, {RASNZ, PO Box 3181, Wellington 6011, New Zealand} \\ 
$^{8}$Variable Stars South, Congarinni Observatory, Congarinni, NSW, 2447, Australia\\
$^{9}$Nielsen, 675 6th Ave, New York, NY 10011, USA\\
$^{10}${{Dept.\ of} Physical Science \& Engineering, Harper College, 1200 W Algonquin Rd, Palatine, {Illinois} 60067, USA}\\
$^{11}$Physics Department, University of Winnipeg, 515 Portage Avenue, Winnipeg R3B 2E9, Canada\\
}

\date{Accepted 2023 November 10. Received 2023 November 10; in original form 2023 September 12}

\pubyear{2023}

\begin{document}
\label{firstpage}
\pagerange{\pageref{firstpage}--\pageref{lastpage}}
\maketitle

\begin{abstract}
New photometry, including TESS data, have been combined with recent spectroscopic observations of the Orion Ib pulsating triple-star system VV Ori.  This yields a revised set of absolute parameters with increased precision. Two different programs were utilized for the light curve analysis, with results in predictably close agreement. The agreement promotes confidence in the analysis procedures. The spectra were analysed using the {\sc FDBinary} program. The main parameters are as follows: $M_1 =  11.6 \pm 0.14$ and $M_2 = 4.8 \pm 0.06$ (M$_\odot$). We  estimate an approximate mass of the wide companion  as $M_3 = 2.0 \pm 0.3$  M$_\odot$.  Similarly, $R_{1} = 5.11 \pm 0.03$, $R_2 =  2.51 \pm 0.02$, $R_3 =  1.8 \pm 0.1$ (R$_\odot$); $T_{\rm e 1} = 26600 \pm 300$,  $T_{\rm e 2} = 16300 \pm 400$ and $T_{\rm e 3} = 10000 \pm 1000$ (K).  The close binary's orbital separation is $a= 13.91$  (R$_\odot$); its age is $8 \pm 2$ (Myr) and its photometric distance is $396 \pm 7$ pc.  The primary's $\beta$ Cep type oscillations support these properties and confirm our understanding of its evolutionary status.   Examination of the well-defined $\lambda$6678 He I profiles reveals the primary to have  a significantly low  projected rotation: some 80\% of the synchronous value. This can be explained on the basis of the precession of an unaligned spin axis.  This proposal can resolve also observed variations of the apparent inclination and  address other longer-term irregularities of the system reported in the literature. This topic invites further observations and follow-up theoretical study of the dynamics of this intriguing young multiple star.

\end{abstract}




\section{Introduction}
\label{section:introduction}  

\begin{figure}
    \begin{center}
        \includegraphics[height=8cm]{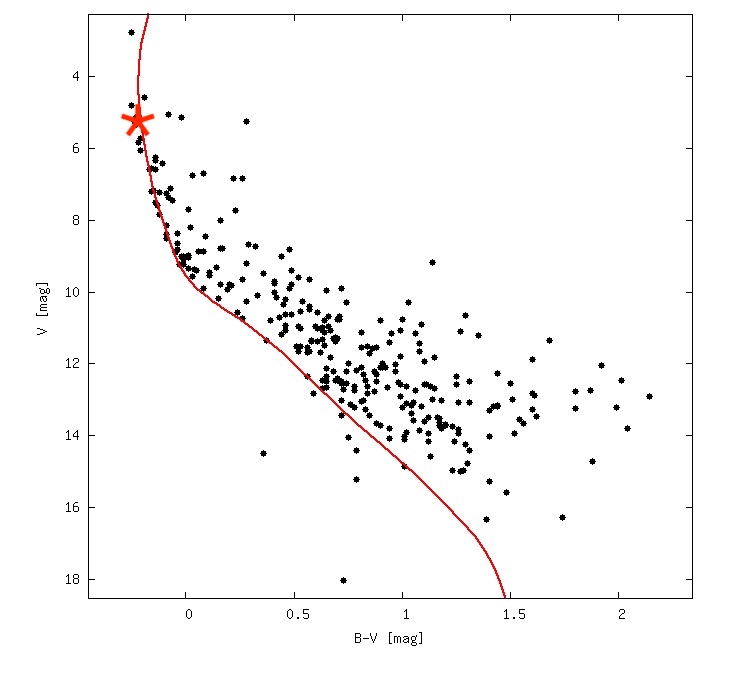}
        \caption{{\sc webda} colour-magnitude diagram based on the data of \protect \cite{Warren_1977}. The Padova isochrone for log (age) = 6.78 y is shown with solar composition, reddening $E(B-V) = 0.05$ and distance modulus 8.16.  VV Ori is marked by a red coloured star.}
        \label{fig:Orion1bcm} 
    \end{center}
\end{figure}

VV Ori is a very bright ($V \sim 5.4$, $M_{V} \sim -2.8$ mag) and well-known eclipsing binary in the Belt, or $\epsilon$ Ori, grouping of the Orion 1b association {\citep{Blauuw_1964, Wright_2020}.}  Superficially, it resembles the system V Pup, with its period of $\sim$1.5 d and B1 + B3-5 type components (cf. \citealt{Budding_2021}). However, numerous studies of the close pair in VV Ori have found a detached arrangement of young early-type main-sequence stars (cf.\ Fig~\ref{fig:Orion1bcm}), unlike the semi-detached configuration of V Pup. As the binary's orbital plane is close to the line of sight, a succession of complete (or nearly so) eclipses is observed, allowing confidence about the determination of fitting function parameters including those relating to the distribution of brightness over the stellar surfaces. 

Massive young stars like those in VV Ori are frequently found in groups of multiple star systems \citep{Sana_2012}. Such star formation regions are thought to have an important role in determining the long-term behaviour of galaxies \citep{Langer_2012, Zucker_2022}. A striking recent discovery about VV Ori, enabled by the long duration and high quality of TESS photometry \citep{Ricker_2014}, is the existence of low-level $\beta$ Cep type pulsations that may be associated with the primary star  \citep{Southworth_2021}.

It is well known that essential properties of stars can be determined from combining the analysis of photometric data with spectroscopic data  --- the `eclipse method'. We follow that approach for VV Ori. This  has been more recently enhanced by the understanding  of stellar pulsational behaviour, or asteroseismology  (\citealt{Aerts_2010}; \citealt{Murphy_2018}, \citealt{Bowman_2020}). A decomposition of the pulsations into constituent oscillation modes will furnish direct information on the stellar structure, particularly if constraints are imposed by membership of a co-evolutionary group of nearby stars \citep{Lampens_2006}. 
 
With regard to early (B type) stellar pulsators, those of the $\beta$ Cep \citep{Lesh_1978} and slowly-pulsating (SPB) kinds \citep{Waelkens_1991} are well known. $\beta$ Cep variables show low-order gravity (g) and pressure (p) oscillation modes, with typical brightness variations of order 0.1 mag.  They are associated with pulsation periods in the range 2 to 6 hr \citep{Stankov_2005} and masses of around  8 to 15 M$_{\odot}$. SPB pulsations have quite longer periods, generally a few days, and are found in the lower mass range of 3 to 9 M$_{\odot}$. The variability is believed to result from high-order g-modes.  More recent space-based observations, however, have uncovered both p- and g-type low-amplitude pulsations in many massive stars  outside the foregoing ranges (\citealt{Pedersen_2019}; \citealt{Burssens_2020}).

Combining regular light and radial velocity (RV) curve analyses with astroseismology, especially with recently increased data accuracy, permits stellar masses and radii to be more confidently specified (\citealt{Ratajczak_2017}; \citealt{Southworth+2020,Southworth_2021}; \citealt{Salmon_2022}).   That being said, stellar  oscillations in massive eclipsing binaries have only been definitely established in relatively few cases  so far (e.g., \citealt{Clausen_1996}, \citealt{SouthworthBowman}, \citealt{Erdem_2022}). Stellar structural parameters may  also be checked from eccentric close binary systems through non-Keplerian (apsidal) motion that relates to the scale of their proximity effects (\citealt{Sterne_1939}; \citealt{Russell_1942};  \citealt{Kopal_1959}). For more recent discussion, see also \citet{Welsh_2011}; \citet{Hambleton_2013}; \citet{Feiden_2015}; \citet{Bowman_2019}; \citet{Handler_2020}; \citet{Kurtz_2020} and \citet{Fuller_2020}. A review of the use of space-based photometry for binary star science can be found in \citet{Southworth_Univ}.

VV Ori contains components of spectral types B1 V and mid-B V in an essentially circular orbit of period 1.485 d. Its eclipsing nature was discovered in 1903 \citep{Barr_1905}.   The early history of its study was summarised by \cite{Wood_1946} and \cite{Duerbeck_1975}.  Two of the more recent published investigations of the system parameters are those of \cite{Sarma_1995} and  \cite{Terrell_2007}.

Multiplicity, beyond the two eclipsing stars, has been discussed in a number of studies over the years. Excess scatter in early RV data led to cautious suggestions of a third body with an orbital period of $\sim$120 d (\citealt{Daniel_1915}; \citealt{Struve_1949}), a proposal that has been disfavoured in later papers (\citealt{Terrell_2007}, \citealt{Van_Hamme_2007}).  These latter papers cast doubt on the evidence for a close third body, at least regarding the RV analyses, though various photometric studies (\citealt{Budding_1980}; \citealt{Chambliss_1983}; \citealt{Van_Hamme_2007}) admitted a small third light contribution. However, \cite{Horch_2017} have resolved a companion at an angular separation of 0.23 arcsec using speckle interferometry. The magnitude differences between this companion and the binary system are 3.88 mag at 692 nm and 3.43 mag at 880 nm.  Applying a linear photometric gradient approximation \citep{Golay_1974} would  yield a magnitude difference of about 4.45 in the V range, or about 0.017 of the system's light, which is essentially the same value found by \cite{Chambliss_1983}. At a distance of $441 \pm 22$ pc \citep{Gaia23}, the separation corresponds to a projected distance of $\sim$87 au, and thus a minimum orbital period of around 200 yr. This resolved companion cannot, therefore, be directly  responsible for the putative 120 d orbital variations.

Recently, \cite{Southworth_2021} noted that the succession of well-observed minima, over the years, has shown a transition between complete and partial eclipses. This can be associated with a systematic variation of the apparent inclination of the binary orbit to the line of sight.  That there may be some variations of  shorter period than that of the wide orbit is thus still an open question.

A consensus from radial velocity studies (\citealt{Daniel_1915}; \citealt{Struve_1949}; \citealt{Beltrami_1969}; \citealt{Duerbeck_1975}; and \citealt{Popper_1993}) presented the system as mostly single-lined, though \cite{Beltrami_1969} were able to model blended line profiles to derive a velocity separation and thence a mass ratio of $\sim$0.45; similar to that of \cite{Duerbeck_1975}.  

In this study, we bring new evidence to bear on outstanding uncertainties about the system, regarded as important for fixing the properties of massive early-type stars (\citealt{Eaton_1975}; \citealt{Popper_1993}; \citealt{Terrell_2007}; \citealt{Van_Hamme_2007}; \citealt{Southworth_2021}). We examine new high quality photometry from the TESS satellite, and high-dispersion spectroscopy from the {\sc fies} spectrograph of the 2.56-m Nordic Optical Telescope, Tenerife, Spain; as well as the {\sc hercules} spectrograph of the University of Canterbury Mt John Observatory, New Zealand. These new data yield important revisions in the physical properties of the component stars. We use up-to-date analytical tools to bear on the data analysis,   and check on evolutionary models, keeping in mind additional evidence relating to the Orion OB 1b  membership and the degree of multiplicity of VV Ori. 


\section{Photometry}
\label{section:photometry}

Table~\ref{tab:history} presents a summary of the main parameters derived in previous photometric studies of VV Ori light curves (LCs) at visual (V) wavelengths.  Standard notation is used, i.e.\ $L_n$ for fractional luminosities (where $n$ indicates a given star in the system), $r_n$ for relative radii, $i$ for the orbital inclination, and $u_n$ for linear limb-darkening coefficients. The sources of these parameters are as follows: Du-75 = \cite{Duerbeck_1975}; Na-81 = \cite{Budding_1980}; Ch-84 = \cite{Chambliss_1984}; Sa-95 = \cite{Sarma_1995}; Te-07 = \cite{Terrell_2007}; So-21 = \cite{Southworth_2021}. Some variations of parameter values can be noticed across Table~\ref{tab:history}, but these are generally in keeping with the listed uncertainties.  This table is not intended to be exhaustive, but it gives an impression of reference quantities derived from data analysis over the last several decades, keeping in mind concomitant increases of data accuracy.

\begin{table*}
\caption{Reference parameters for historic photometric model fits.} 
\begin{tabular}{lcccccc}
\hline  
\multicolumn{1}{c}{Parameter}  & \multicolumn{1}{c}{Du-75} & 
\multicolumn{1}{c}{Na-80} & \multicolumn{1}{c}{Ch-84} & \multicolumn{1}{c}{Sa-95} &  \multicolumn{1}{c}{Te-07}  & \multicolumn{1}{c}{So-21}\\ 
\hline 
$\lambda$ nm  & 540  & 425  & 530 & 530 & (530)  &   TESS     \\
$M_2/M_1$           & 0.45               & 0.45                 & 0.42               & 0.42             & 0.38               & 0.376                \\     
$L_1$               & 0.878 $\pm$ 0.06   & 0.926 $\pm$ 0.09     & 0.891 $\pm$ 0.10    & 0.878 $\pm$ 0.01  & 0.908 $\pm$ 0.09   & 0.892 $\pm$ 0.07   \\  
$L_2$               & 0.122 $\pm$ 0.04   & 0.069 $\pm$ 0.03     & 0.092 $\pm$ 0.05    & 0.122 $\pm$ 0.01  & 0.098 $\pm$ 0.05   & 0.103 $\pm$ 0.005  \\  
$L_3$               & 0.000 $\pm$ 0.06   & 0.005 $\pm$ 0.03     & 0.017 $\pm$ 0.11    & 0.00 $\pm$ 0.01   & 0.00 $\pm$ 0.07   & 0.005 $\pm$ 0.07   \\ 
$r_1 $              & 0.366 $\pm$ 0.006  & 0.371 $\pm$ 0.01     & 0.363 $\pm$ 0.02    & 0.378$\pm$ 0.02   & 0.369 $\pm$ 0.005  & 0.372 $\pm$ 0.001  \\ 
$r_2$               & 0.180 $\pm$ 0.01   & 0.168 $\pm$ 0.03     & 0.176 $\pm$ 0.05    & 0.179 $\pm$ 0.05  & 0.179 $\pm$ 0.013  & 0.182 $\pm$ 0.003  \\ 
$i$                 & 85.6  $\pm$ 1.0    & 85.8 $\pm$ 1         & 85.6 $\pm$ 2        & 86.1 $\pm$ 5      & 85.9 $\pm$ 0.5     & 78.3 $\pm$ 0.5     \\ 
$T_h$ (K)           & 25000              & 25000                &     25000           &      25000        &     26199          & 26200              \\
$T_c$ (K)           & 15000              & 16000                &      15700          &       15500       &      16073         & 16100              \\  
$u_1$               & 0.07               & 0.37                 & 0.28                & 0.33              & --                 &    0.63 (bol)       \\   
$u_2$               & 0.45               & 0.45                 & 0.32                & 0.37              & --                 &     0.71 (bol)            \\ 
\hline
\label{tab:history}
\end{tabular}
\end{table*}

\begin{figure}
    \begin{center}
        \includegraphics[width=\linewidth]{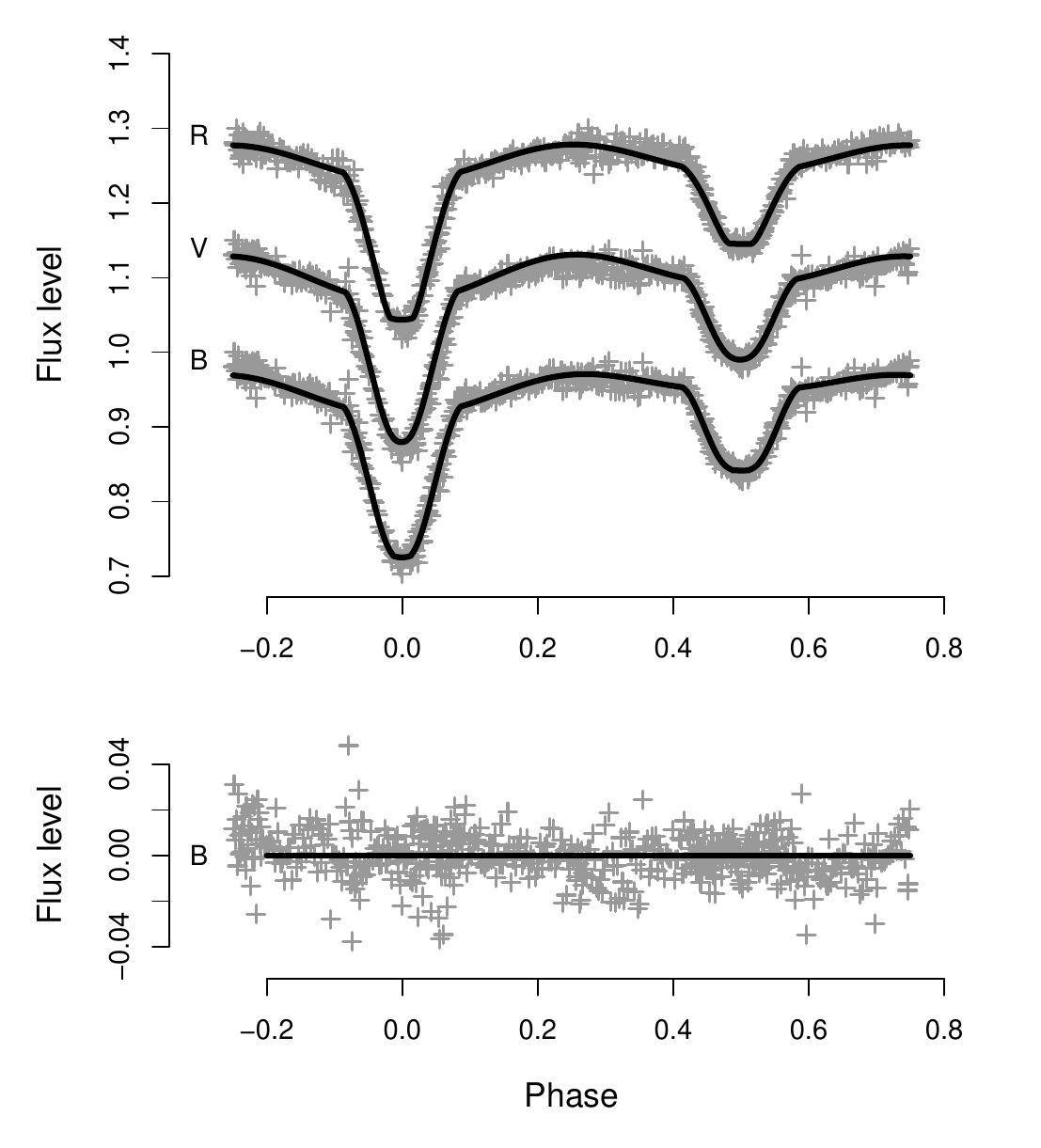}
        \caption{{\sc WinFitter} model fitted to recent BVR light curves from the Congarinni  Observatory. The V and R light curves have been vertically offset by 0.15 and 0.30 to allow presentation on a single chart.} 
        \label{fig:Orimarkbvr} 
    \end{center}
\end{figure}

Further multi-colour photometry of VV Orionis (see Table~\ref{tab:congarinni}) was carried out over ten nights between December 2018 and February 2019 from the Congarinni Observatory, NSW, Australia (152$^\circ$ 52$^\prime$ E, 30$^\circ$ 44$^\prime$ S, 20 metres above mean sea level). Images were captured with an ATIK$^{\rm TM}$ One 6.0 CCD camera equipped with Johnson-Cousins BVR filters attached to an 80mm f6 refractor, which, given the brightness of the stars, was stopped down to 50mm aperture.  MaxIm DL$^{\rm TM}$ software was used for image handling, calibration and aperture photometry. HD 36779 was used as the main comparison star.  Its magnitude and colours were determined as V = 6.223, B -- V 
= --0.165 and V -- R = --0.067, in close agreement with the Johnson 11-colour catalogue  \citep{Ducati_2002}.  The derived light curves are shown in Fig.~\ref{fig:Orimarkbvr}. Optimal parameters from {\sc WinFitter} modelling (cf.\ \citealt{Rhodes_2022}) are given in Table~\ref{tab:congarinni_fit}. Using the results given in Table~\ref{tab:congarinni} as a guide, we derived BVR magnitudes of the three identified photometric  components of VV  Ori as: 5.23, 8.00, 9.95 (B); 5.41, 8.02, 9.79 (V); 5.65, 8.09, 9.61 (R).  Taking into account local variations of the mean reddening ($\sim$0.09, according to  \citeauthor{Warren_1978}, \citeyear{Warren_1978})  and errors of measurement, these figures are in fair accord with the assigned early and mid-B main sequence types for the close pair attended by a cooler, A-F type, companion. 

\begin{table}
\caption{Summary of Congarinni BVR photometry of VV Ori, along with uncertainties, given as standard deviations. 
}
\begin{tabular}{lccccccc}
\hline  
\multicolumn{1}{c}{Date}  & \multicolumn{1}{c}{Phase} & 
\multicolumn{1}{c}{V} & \multicolumn{1}{c}{stdev} & \multicolumn{1}{c}{B -- V} &  \multicolumn{1}{c}{stdev}  & \multicolumn{1}{c}{V -- R } & \multicolumn{1}{c}{stdev} \\ 
\hline 
 19\, 01\, 26 & 0.0  & 5.627 & 0.034 & --0.140 &  0.042& --0.113 & 0.046 \\
 18\, 12\, 25 & 0.5  & 5.465 & 0.017 & --0.157 &  0.024& --0.114 & 0.021 \\
 18\, 12\, 28 & 0.5  & 5.467 & 0.016 & --0.166 &  0.021& --0.113 & 0.023 \\
 18\, 12\, 29 & 0.25 & 5.308 & 0.022 & --0.157 &  0.029& --0.114 & 0.034 \\
 19\, 01\, 25 & 0.25 & 5.314 & 0.022 & --0.157 &  0.032& --0.104 & 0.035 \\
 18\, 12\, 29 & 0.75 & 5.281 & 0.020 & --0.158 &  0.028& --0.125 & 0.021 \\
\hline
\label{tab:congarinni}
\end{tabular}
\end{table}

\begin{table}
\caption{Parameters for model fits to the BVR photometry shown in Fig~2.  Adopted parameters are listed below the 3rd row horizontal line. The geometric elements $r_1$, $r_2$ and $i$, are  weighted averages from initial fittings at the separate wavelengths. The value of $L_3$ in the B filter, being expectably less than its uncertainty as an optimal fitting result, is here a fixed hyperparameter.}
\label{tab:congarinni_fit}
\begin{tabular}{llll}
\hline  
\multicolumn{1}{c}{Parameter}  & \multicolumn{1}{l}{B} & 
\multicolumn{1}{l}{V} & \multicolumn{1}{l}{R}\\ 
\hline 
$L_1$               & 0.926 $\pm$ 0.012  & 0.901 $\pm$  0.013  & 0.894 $\pm$ 0.013   \\
$L_2$               & 0.072 $\pm$ 0.003  & 0.082 $\pm$  0.003  & 0.085 $\pm$ 0.003   \\ 
$L_3$               & 0.012            
& 0.016 $\pm$  0.014  & 0.021 $\pm$ 0.013   \\ 
\hline
$r_1 $ (mean)       & \multicolumn{3}{l}{0.359 $\pm$ 0.008}     \\ 
$r_2$ (mean)        & \multicolumn{3}{l}{0.171 $\pm$ 0.004}     \\ 
$i$ (deg, mean)     &  \multicolumn{3}{l}{80.0 $\pm$ 0.8}      \\
$M_2/M_1$           &  \multicolumn{3}{l}{0.42}     \\  
$T_h$ (K)           & \multicolumn{3}{l}{26600}                         \\
$T_c$ (K)           & \multicolumn{3}{l}{16250}                         \\  
$u_1$               & 0.28               & 0.25                 & 0.21            \\   
$u_2$               & 0.29               & 0.29                 & 0.24          \\ 
\hline
$\chi^2/\nu$        & 1.01               & 1.05                 & 0.90           \\ 
$\Delta l $         & 0.014               & 0.012                 & 0.010           \\ 
\hline
\end{tabular}
\end{table}

\subsection{Light curve analysis}
\label{section:light_curve_analysis}

\begin{table}
\caption{ Summary of the parameters for the WD fittings of  the TESS sector 6, phase-binned light curve of VV Ori. Detailed descriptions of the control parameters can be found in the WD code user guide \protect\citep{Wilson_2004}. The uncertainties have been determined via comparisons  through a wide range of model fits.
\label{tab:WD_fittings_sunmmary}}
\begin{center}
\begin{tabular}{lrr}
\hline
\multicolumn{1}{l}{Parameter} & \multicolumn{1}{r}{WD2004 name} & 
\multicolumn{1}{r}{Value} \\  \hline
{\em Control \& fixed parameters:}          &               &                       \\
WD2004 operation mode                       &   MODE        & 0                     \\
Treatment of reflection                     &   MREF        & 1                     \\
Number of reflections                       &   NREF        & 1                     \\
LD law                                      &   LD          & 2 (logarithmic)       \\
Numerical grid size (normal)                & N1, N2        & 60                    \\
Numerical grid size (coarse)                & N1L, N2L      & 60                    \\
\\
{\em Fixed parameters:}                     &               &                       \\
Mass ratio                                  & RM            & 0.418                 \\
Orbital eccentricity                        & E             & 0.0                   \\
$T_{\rm e}$ of primary (K)                  & TAVH          & 26,660                \\
$T_{\rm e}$ of secondary (K)                & TAVH          & 16,250                \\
Bolometric albedos                          & ALB1, ALB2    & 1.0,  1.0             \\
Rotation rates                              & F1, F2        & 1.0, 1.0              \\
Gravity darkening                           & GR1, GR2      & 1.0, 1.0              \\
Logarithmic LD coefficients                 & Y1A, Y2A      & 0.217, 0.205          \\
\\
{\em Fitted parameters:}                    &               &                       \\
Primary potential                           & PHSV          & 3.220 ± 0.026         \\
Secondary potential                         & PHSV          & 3.637 ± 0.023         \\
Orbital inclination (deg)                   & XINCL         & 79.77 ± 0.19          \\
Primary light contribution                  & HLUM          & 10.48 ± 0.10          \\
Secondary light contribution                & CLUM          & 1.215 ± 0.012         \\
Third light                                 & EL3           & 0.071 ± 0.008         \\
Primary linear LD coefficient               & X1A           & 0.359 ± 0.049         \\
Secondary linear LD coefficient             & X2A           & 0.22 ± 0.15           \\
Fractional radius of primary                &               & 0.3656 ± 0.0019       \\
Fractional radius of secondary              &               & 0.1794 ± 0.0015       \\
\hline
\end{tabular}
\end{center}
\end{table}

\begin{figure}
\includegraphics[width=\columnwidth]{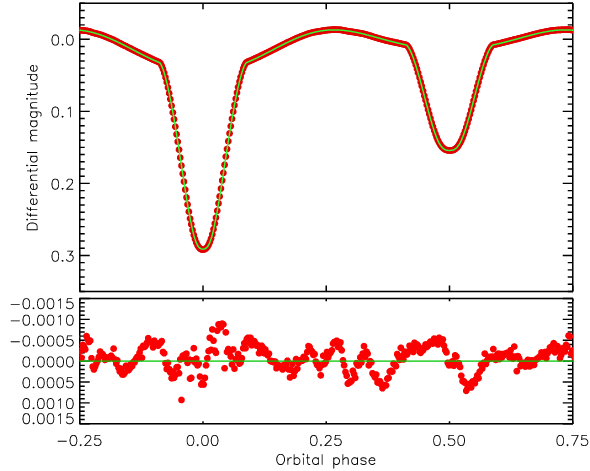} \\
\caption{\label{fig:wd}  The adopted optimal WD model (green line) to the TESS sector 6 phase-binned light curve of VV \,Ori (red filled circles). The residuals of the fit are plotted in the lower panel using a greatly enlarged y-axis to bring out the detail.}
\end{figure}

We first sought to match the form of the light curve with the well-known  Wilson-Devinney (WD) code \citep{Wilson_1972, Wilson_1979}. We used the 2004 version of the code \citep{Wilson_2004} implemented using the {\sc jktwd} wrapper \citep{Southworth_2011}. This code adopts Roche model geometry for the calculation of the shapes of binary stars (cf.\ \citealt{Kopal_1959}; ch.\ 3), although it becomes computationally expensive to perform on the original complete set of TESS observations. We therefore took the data from sector 6, converted them to orbital phase, and averaged them into 400 equally-spaced phase bins.

These data were then fitted using WD to obtain the simplest initial model that matched them well. This, our default solution, was obtained by fitting the following model parameters: surface potentials of the two stars, orbital inclination, phase of the primary mid-eclipse, and light contributions of both eclipsing stars together with a possible third light. {The third light is expressed as a fraction of the total brightness of the system at phase 0.25. The light contributions from the two stars are expressed in the WD code on a different flux scale, and add to $4\pi$ minus the third light for zero differential magnitude.} The fitting was performed in mode 0 (see Table~\ref{tab:WD_fittings_sunmmary}), where the effective temperature ($T_{\rm e}$) values of the two stars are fixed and their light contributions matched directly. We therefore fixed the $T_{\rm e}$ values and mass ratio at the adopted spectroscopic values, presented in Section~\ref{section:spectrometry}, and assumed a circular orbit. We also assumed, for the present purpose, synchronous rotation, as well as albedo parameters and gravity darkening exponents set to unity. The limb darkening (LD) effect was accounted for using a logarithmic law with one coefficient adjustable and the other fixed, for both stars. Maximum numerical precision was used, by setting the quantities N1, N2, N1L and N2L to 60. For the effective wavelength of observation, the Cousins $R$ band was adopted as it is the closest  available approximation to the TESS passband. The `simple'  option for the reflection effect was chosen.

%

With these settings in place, we were able to obtain a satisfactory fit to the phase-binned data as shown in Fig.~\ref{fig:wd}.  The residuals are shown on a larger scale to demonstrate the presence of  structure remaining in the phase-binned data. Whilst the differences between observed and calculated LCs are relatively small (an r.m.s.\ of 0.3 mmag), the remaining pulsational signature means the residuals are dominated by red noise. Using fitting procedures that allowed for orbital eccentricity, we conclude that the quantity $e \cos \omega$, where $e$ is the eccentricity and $\omega$ the longitude of periastron, must be smaller than 0.001. A summary of the fitted and fixed parameters is given in Table~\ref{tab:WD_fittings_sunmmary}. The fractional stellar radii specified in Table~\ref{tab:WD_fittings_sunmmary} are volume-equivalent values.  Table~3-2 in \citeauthor{Kopal_1959}'s (\citeyear{Kopal_1959}) book shows these radii to be one or two percent larger than the corresponding unperturbed radii ($r_0$) at the same mass ratio.

Determination of the uncertainties of the fitted parameters is not trivial.  We follow the approach of \citet{Southworth_2020} on this. The uncertainties derive almost entirely from the model limitations, even taking into account the residual pulsation signature, because the precision of the data is extremely high, showing up systematic effects that are not in the model. We therefore ran WD with selected differences in parameter values to determine their effects on the resulting fits, as follows:
\begin{itemize}
    \item{changing the mass ratio by the uncertainty in the spectroscopic value;}
    \item{fixing $e \sin \omega = 0.005$, instead of assuming a circular orbit;}
    \item{decreasing the rotation parameters for the stars by 10\%;}
    \item{changing the albedos by 10\%;}
    \item{changing the gravity darkening exponents by 10\%; }
    \item{lowering the N1, N2 values from 60 to 55 in steps of 1;}
    \item{fitting for $T_{\rm e}$ in mode 2, instead of fitting for HLUM and CLUM in mode 0;}
    \item{using the more detailed reflection model;}
    \item{specifying the square-root instead of the logarithmic limb darkening law} 
    \item{using the Cousins $I$ filter for the TESS pass-band;} 
    \item{fixing third light to be zero;} 
    \item{ changing the number of phase bins from 400, to 300 or 500.}
\end{itemize}

The net result of these tests was a compilation of values for the fitted parameters; each group associated with a different model run. We rejected all runs where the fit was significantly worse (more than 0.35 mmag r.m.s.\ in the residuals), and added all the parameter differences in quadrature to determine the uncertainty for each fitted parameter. These are given in Table~\ref{tab:WD_fittings_sunmmary}.  The largest contributions to these uncertainties were found to come from the effects of the adopted rotation, albedo and eccentricity parameters. The derived uncertainty percentages are nonetheless still small, and the fitted parameters have highly consistent values.

\begin{figure}
\includegraphics[width=\linewidth]{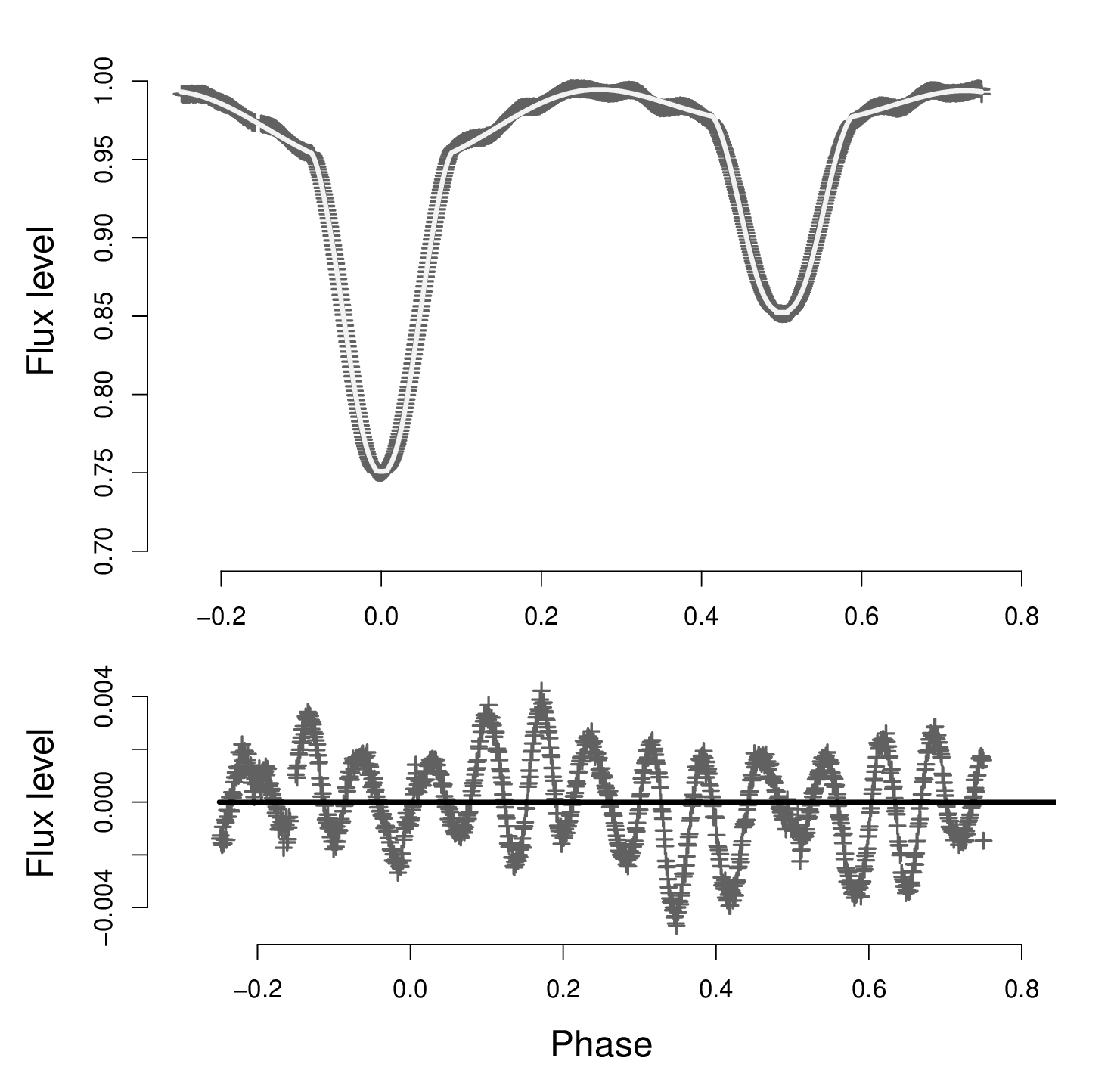}
\caption{The upper panel presents an optimal model LC to  normalized SAP flux measures from TESS for the orbit 13 of Sector 6, TBJD 1487.38930345 -- 1488.87467768, 1070 data points were taken from the original source with no binning.
The lower panel reveals the $\beta$ Cep type behaviour in the residuals.   
\label{fig:tess_lc_fit}
}
\end{figure}

\begin{table}
\caption{Optimal WinFitter parametrization of TESS photometry from Sectors 6 and 32, modelling the entire sectors. The data-sets were phased by the orbital period, thus averaging out the pulsations. The column titled `Figure 4' lists the parameter estimates for the data period plotted in Figure~\protect\ref{fig:tess_lc_fit}.  The mass ratio, stellar temperatures, and limb darkenings are held
constant across these fits. 
} 
\begin{tabular}{lccc}
\hline  
Parameter           & Sector 6                 & Sector 32              & Figure 4              \\
\hline 
$L_1$               & 0.847 $\pm$ 0.010         & $ 0.875 \pm 0.006$    & $0.860 \pm 0.013$     \\
$L_2$               & 0.098 $\pm$ 0.003         & $ 0.096 \pm 0.002$    & $0.100 \pm 0.004$     \\ 
$L_3$               & 0.055 $\pm$ 0.005         & $ 0.008 \pm 0.003$    & $0.040 \pm 0.005$     \\ 
$r_1$               & 0.371 $\pm$ 0.002  & $ 0.370 \pm 0.002$    & $0.368 \pm 0.002$     \\
$r_2$               & 0.183 $\pm$ 0.003  & $ 0.184 \pm 0.003$    & $0.181 \pm 0.001$     \\
$i$ (deg)            & 78.2$\pm$ 0.4    & $ 78.0 \pm 0.4$       & $79.0 \pm 0.5$     \\

\hline
$M_2/M_1$           & \multicolumn{3}{c}{0.48}                          \\ 
$T_h$ (K)           & \multicolumn{3}{c}{26600}                         \\
$T_c$ (K)           & \multicolumn{3}{c}{16250}                         \\  
$u_1$               & \multicolumn{3}{c}{0.18}                          \\  
$u_2$               & \multicolumn{3}{c}{0.20}                          \\ 
\hline
$\chi^2/\nu$        & 1.14                      & 0.97                  & 0.90                  \\ 
$\Delta l $         & 0.0013                     & 0.002                 & 0.0025                \\ 
\hline
\label{tab:tess_fit}
\end{tabular}
\end{table}

For an independent assessment of the LC we used the program WinFitter (WF), discussed in \citet{Budding_2022} chapter 7.  The algebraic form of its fitting function means that parameter space can be searched rapidly and thoroughly using relatively large data-sets.  An example of a LC from TESS Sector 6 is shown in Fig.~\ref{fig:tess_lc_fit} together with an optimal WF model.  The residuals are shown in the lower panel where the $\beta$ Cep oscillations of the primary component are clearly evident.  The adopted light elements for this data set were: 
$$
 {\rm Min \,\, I } =  {\rm TBJD} 1487.389303 + 1.48537742 E. 
$$
LCs from both Sectors 6 and 32 were collected together, phased and binned using this ephemeris. The Sector 6 data were binned with a reduction factor of 14:1 down from 14,847 original data points, while Sector 32 data were binned with the ratio 17:1, from the 17,917 source data. Numerous optimal fitting experiments were performed on these reduced data-sets. Starting values were guided by historic findings. The finally adopted  parameter sets are given in Table~\ref{tab:tess_fit}, and are in reasonable agreement with the WD model parameters.  The separately derived main geometric elements ($r_1$, $r_2$, $i$) of the WD and WF fittings to the TESS data-sets are within the uncertainty estimates of each other.  They concur that the TESS data point to a  distinctly lower inclination than the values cited in Table~\ref{tab:history}. 
 

\subsection{Fitting function with unaligned axes}
\label{section:fitting_function}
It was mentioned above that \cite{Southworth_2021} discussed visible LC effects associated with  an apparent variation of the inclination parameter. The precession of a spin axis that is unaligned to that of the orbit is feasible for a young multiple system 
in a process of angular momentum evolution, and, since the flux from the system is dominated by the primary star, such effects would be mainly associated with  the  behaviour of that star.  In this section we check on the scale of relevant effects and how the WF fitting function may address this topic.

In general, the matching of a theoretical model to a photometric data set (LC) is achieved by optimizing the  agreement of a  fitting function $l(\phi;a_i)$, where $\phi$ is the orbital phase and $a_i$ are a set of parameters whose adjustment is implied by the optimization process. This fitting function can be set in the form:
\begin{equation}
    \label{eq:one}
    l(\phi) = \int_{A_1} J_1 dA_1 +  \int_{A_2} J_2 dA_2  - \int_{A_e} J_e dA_e
\end{equation}
where the suffix 1 refers to the `primary' object, with higher surface temperature, say, having projected surface area $A_1$, while suffix 2 denotes the secondary, projecting area $A_2$ perpendicular to the line of sight. The received flux scales linearly with the integrated locally projected flux $J$ from either source. The suffix $e$ relates to an eclipse: the relevant term is zero when there is no eclipse and maximizes during complete eclipse phases.  Eclipses of the secondary star are covered by the same form, but with the eclipsing and eclipsed roles reversed, $J_e$ being suitably re-assigned. 

Eqn~\ref{eq:one} is reducible to a relatively simple form in the `spherical model' (cf.\ \citealt{Russell_1912}). This reduction suggests that modelling  can be developed to cater for more realistic situations using the  Taylor expansion
\begin{equation}
    \label{eq:taylor_expansion}
    l(\phi) = l_0(\phi) + \Delta l_{A1}(\phi) + \Delta l_{A2}(\phi) - \Delta l_{Ae}(\phi)  \, ,
\end{equation}
where $ l_0(\phi) $ in
Eqn~\ref{eq:taylor_expansion} would be normalized so that the sum of the first two integrals, in the unperturbed, uneclipsed situation, is unity.  In this case, the first term becomes the primary's fractional luminosity $L_1$, the second, correspondingly, $L_2$.

We can adopt, without loss of generality, that suffix 1 refers to the star about to be eclipsed and the zero order LC equation becomes
\begin{equation}
    \label{eq:zero_order}
    l_0(\phi) = 1 - \alpha\{u_1,k,d(r_1, i, \phi)\} L_1  \,  ,
\end{equation}
where $\alpha$ is Russell's (\citeyear{Russell_1912a, Russell_1912b}) light loss function, depending here on  the re-normalized separation $d = \delta/r_1$; $\delta$ being the separation of the two star centres in units of the mean orbital radius on the tangential sky plane, given by 
\begin{equation}
    \label{eq:four}
    \delta^2 = \sin^2\phi \sin^2 i + \cos^2 i .
\end{equation}
Other parameters include the ratio of radii $r_2/r_1$, written as $k$ in Eqn(3), and a limb-darkening coefficient $u_1$.

For a more realistic model we should evaluate the effects of tidal and rotational perturbations of the photosphere, associated with  the proximity of the two stars, at given phases, as well as the `reflection' effects that result from their mutual irradiation.  Calculating the surface distortion involves the radial displacements in the key directions of the line of centres  and the  axis of rotation.   Local areal  projections for any viewing angle are then resolved into their components in the observer's frame of reference. 

Both WD and WF converge to the same approximation for the first order surface perturbation given as
\begin{equation}
    \frac{ \Delta^{\prime} r }{r_0}  = q \sum^{4}_{j = 2} r_0^{j+1} (1 + 2k_j) P_j(\lambda) + n r_o^3(1 - \nu^2)  
    \label{eq:fosp}
\end{equation}
where $r$ is the local stellar radius expressed as a fraction of the orbital separation of the components with mean value $r_0$, and $k_j$ are the well-known structural constants reflecting the distribution of matter through the star.  $\lambda$  is the  direction cosine of the angle between the radius vector $\hat{r}$ and the line of centres, and $\nu$ is the  direction cosine of the angle between $\hat{r}$ and the rotation axis.  The coefficients $k_j$ can be taken from suitable stellar models, e.g.\ \cite{Inlek_2017}, { or set to zero in the Roche model (cf. Eqns 1-11 in Ch.1, and 2-6 in Ch. 3 of \citeauthor{Kopal_1959}, \citeyear{Kopal_1959}).

\begin{figure}
\includegraphics[width=\linewidth]{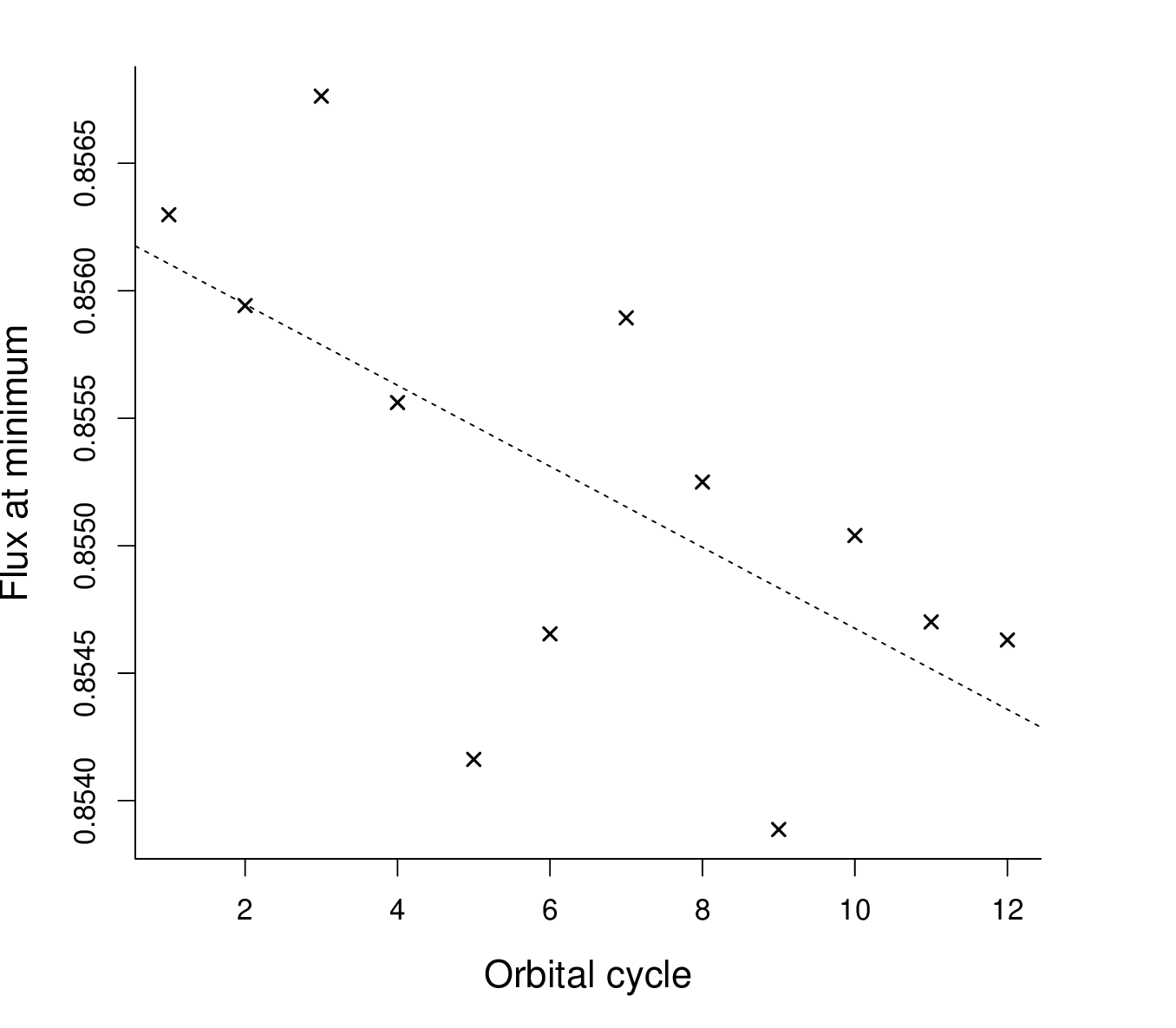}
\caption{A  trend across a flux interval of several mmag can be seen in the succession of TESS monitorings of secondary minima in the Sector 6 data.
\label{fig:tess_lc_secmin}
}
\end{figure}

\begin{figure}
\includegraphics[width=\linewidth]{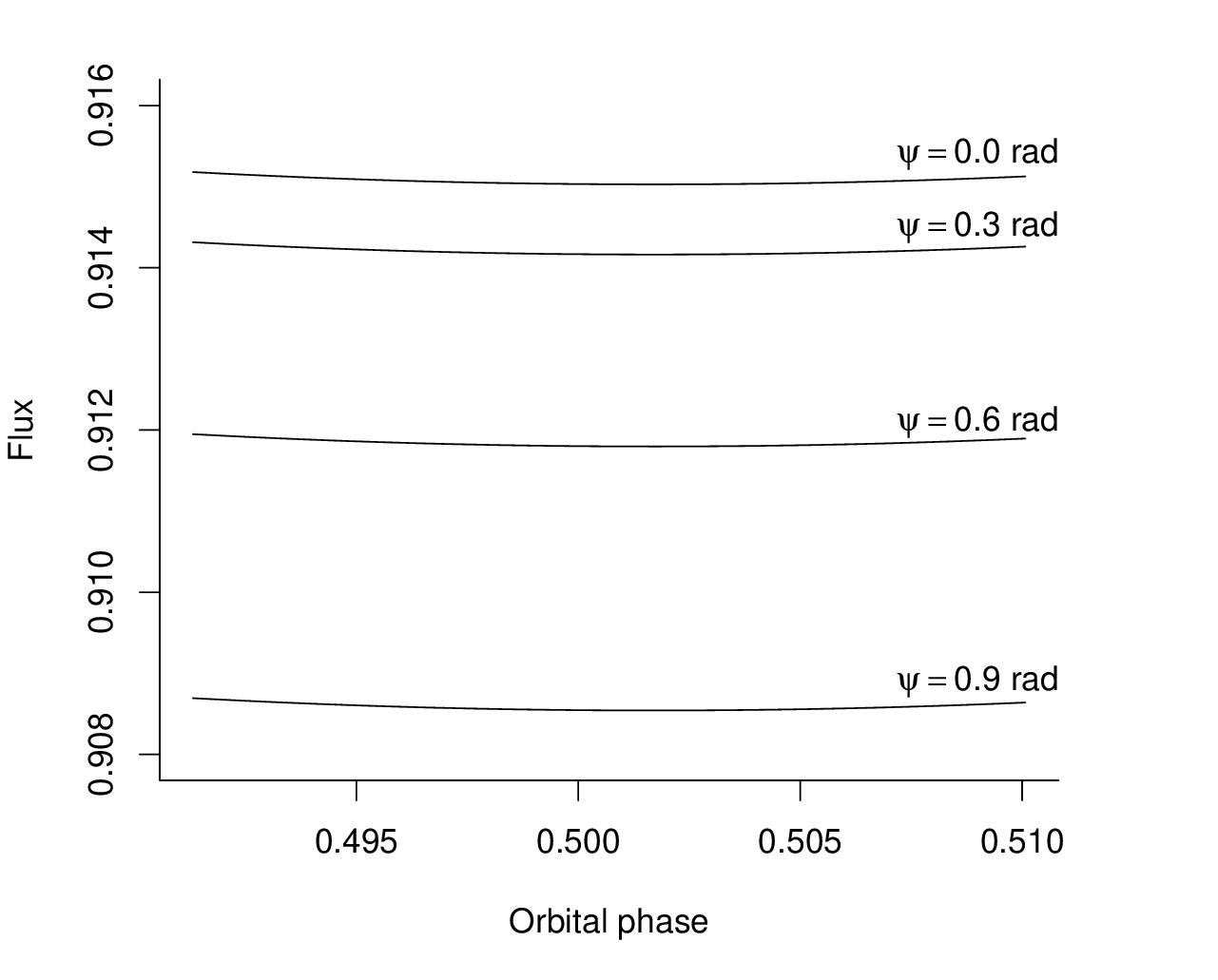}
\caption{The effects of varying the precession angle of the primary star's spin axis, at a 20 deg obliquity, on the relative flux during secondary minimum.
\label{fig:tess_lc_precess}
}
\end{figure}

With the surface perturbations as serial harmonic functions, they are seen (Eqn~\ref{eq:fosp})  to start with terms of order $r^3$, so by the time we reach terms of order $r^6$ the gravitational interaction of tides on tides would be taken into account. WF analysis, in which the mutual effect of perturbations on perturbations is neglected, therefore proceeds up to and including terms of order $r^5$. This implies three separate tidal terms in the fitting function. 

Rotation with a constant angular velocity, as is usually assumed, requires only one source term.  However, location of the rotation axis with respect to the orbit will, in general, involve two additional angular parameters: $\epsilon$ the obliquity and $\psi$ the precession angle.  In the case of aligned rotation and orbit axes, the surface distortions due to rotation, as they appear to a remote observer, involve integrals factored by orientation-dependent direction cosines.  These reduce down to the independent  cosine of the angle between the line of sight and the rotation axis $n_0$. This becomes replaced,  in the unaligned case, by $n_0^{\prime}$, where 
\begin{equation}
    \label{eq:five}
    n_0^{\prime} = n_0  \cos \epsilon + \sin i \sin \epsilon \cos \psi 
\end{equation} 
(cf. Eqn~(1.13) of \citealt{Budding_2022}).

Fig.~\ref{fig:tess_lc_secmin} plots the measured depths of the secondary minima over a succession of LCs from TESS sector 6 with mean epoch TJD 1477.5973. This comes from averaging the values of 13 individual flux measures --- at phase 0.5 and the 6 points on either side. A trend of order 2 mmag becomes apparent over the 18 day time interval.  LC modelling of the partial, but near total, secondary eclipse  shows that this would necessitate a change of inclination significantly greater than the $\sim0.01^{\circ}$  (for 18 d) considered by \cite{Southworth_2021} after inspection of historic LCs. However, the residuals presented in Fig.~\ref{fig:tess_lc_fit} allow an expectation of point-to-point  variations in Fig.~\ref{fig:tess_lc_secmin} of several mmag, so this apparent short-term trend is not separable from effects {\em not} due to inclination changes. On the other hand, LC modelling shows that variations of several mmag in  the central depth of the secondary minimum would result from changes of the apparent inclination, i.e.\ $\arccos ( n_0^{\prime})$, of  order 1$^{\circ}$. Fig.~\ref{fig:tess_lc_precess} shows that this could be accounted for by precessional movement of order $\sim$10$^{\circ}$ on a feasible timescale  of  $\sim$10 yr. 

A further series of TESS LCs  with mean epoch TJD 2186.0373, i.e.\ 708.44 d later, were collected in sector 32.  Unfortunately, this nearly 2 yr time baseline produces only a 0.4$^{\circ}$ shift in the inclination at the rate presented by \citeauthor{Southworth_2021}, so comparable to the uncertainty in the estimated inclination values of Tables 4 and 5.   The mean relative fluxes  at mid-secondary minimum from the two sectors are 0.8549$\pm$0.0015 and 0.8551$\pm$0.0014, which does not yet provide convincing support for a secular change in the relative depth of the secondary minimum over the two years separating the two TESS sectors involved.


\subsection{Evaluation of changes between TESS sectors} 
\label{section:tessvary}

\begin{figure}
\includegraphics[width=\linewidth]{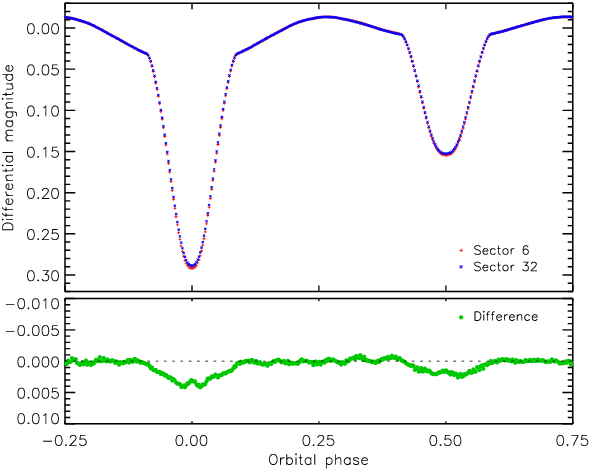}
\caption{\label{fig:tessvary} Phase-binned versions of sectors 6 and 32  TESS observations (top) and their difference on a magnified scale (bottom).}
\end{figure}

The hypothesis of orbital evolution associated with  a third body, or some other cause, leads naturally to the suggestion that the light curve may change between the two TESS sectors. \cite{Southworth_2021} found a change in orbital inclination which in turn may change the eclipse depths. We therefore searched for these effects.

We first fitted the two TESS sectors simultaneously to obtain an orbital ephemeris which precisely matched these datasets without influence from other data. We then phase-binned each TESS sector into 400 bins (as done above). A plot of these sectors and the difference between them can be found in Fig.\,\ref{fig:tessvary}. A decrease in the eclipse depths is clear.

We then fitted both datasets using the Wilson-Devinney code (see Section\,\ref{section:light_curve_analysis}). For each fit we adopted one approach to modelling the data-sets, and analysed them separately. We performed a range of fits varying the numerical precision of the code, the treatment of rotation, albedo and limb darkening. The differences in orbital inclination across all fits averaged at close to 0.17$^\circ$ (mean) with a standard deviation of 0.08$^\circ$. We therefore find a change in inclination to a significance level of 2$\sigma$ over a time period of 706\,d. The LC clearly changes, but we are not yet able to quantify this with sufficient statistical significance. It is unfortunate that no further observations of VV\,Ori by TESS are currently scheduled.

An alternative explanation of the change in eclipse depth is imperfections in the data reduction. An obvious possibility is an error in the background subtraction which would shift the light curve up or down by a fixed amount of flux. In this case we would expect to see the difference between the TESS sectors (Fig.\,\ref{fig:tessvary}) appear as a scaled-down version of the light curve. That is not the case: the differences between sectors occur only during the eclipses. This supports the possibility of changing apparent inclination in the eclipsing pair in the VV\,Ori system.


\subsection{Frequency analysis of TESS data} 
\label{section:frequency_analysis}

\begin{figure} 
\includegraphics[width=\linewidth]{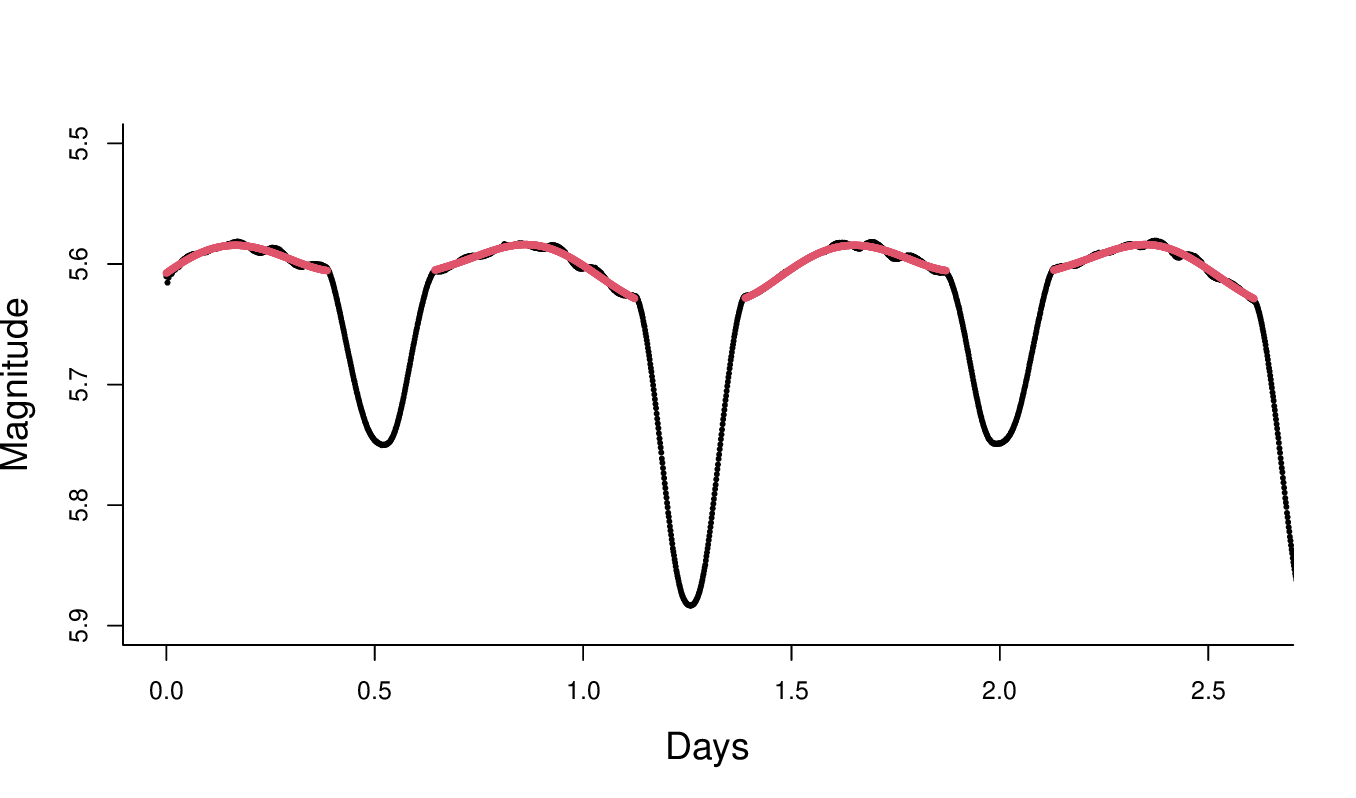}
\caption{The first 2.5 days of TESS Simple Aperture Photometry (SAP)  from Sector 32. Out-of-eclipse regions used for frequency analysis are shown in red.
\label{fig:tess5res1}}
\end{figure}

\begin{figure} 
\includegraphics[width=\linewidth]{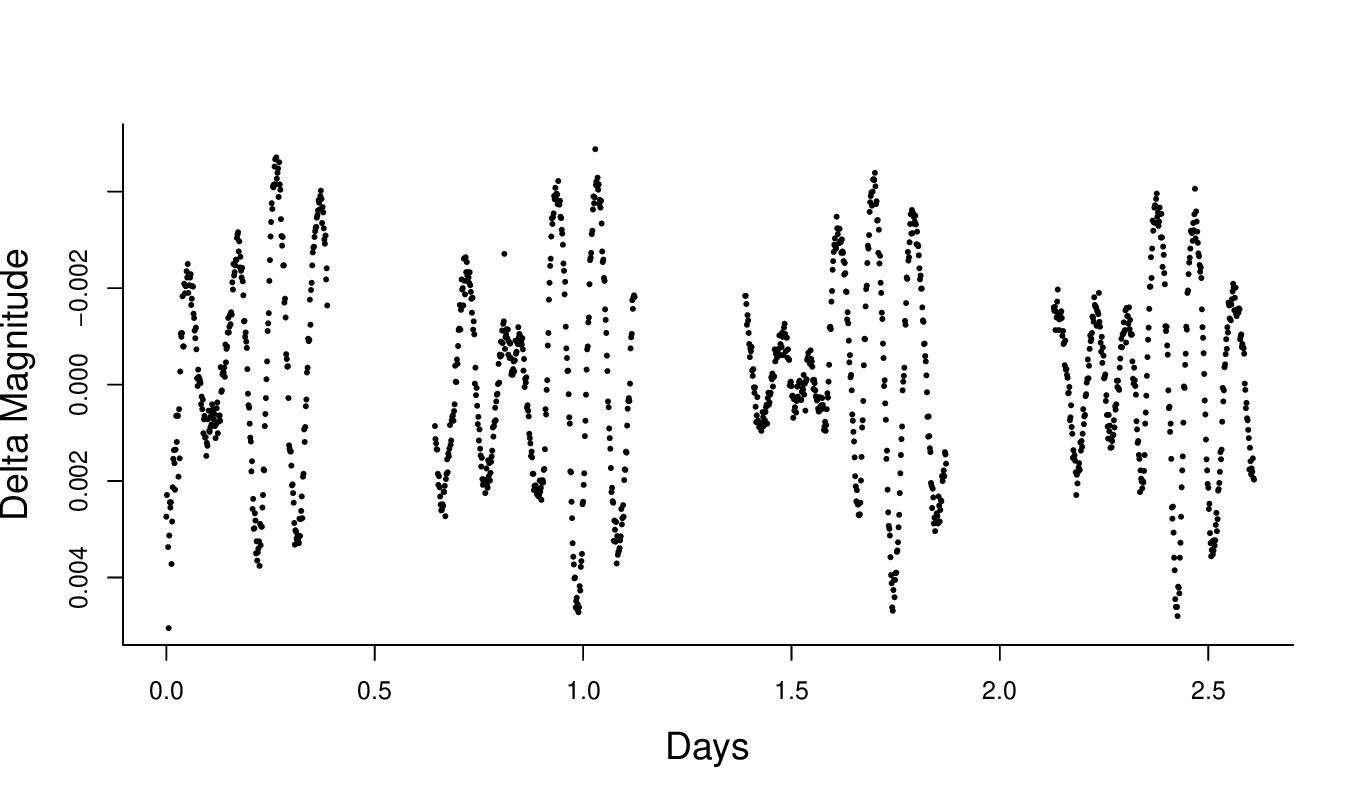}
    \caption{Residual pulsations  in the inter-minimum  phases after removing the eclipsing binary LC. 
\label{fig:tess5res1b}}
\end{figure}

\begin{figure*}
\includegraphics[width=\linewidth]{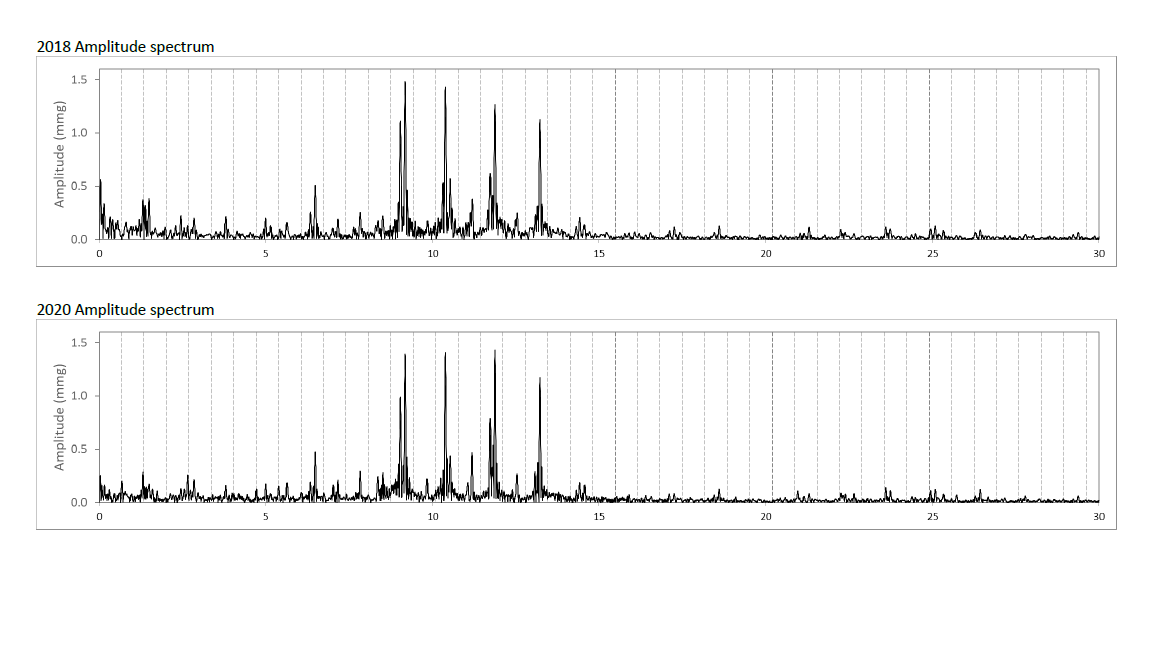}
\vspace*{-24mm}
    \caption{Frequency spectra of VV Ori using {\sc period04} on the residuals from the 2018 and 2020 binary LC fittings to the TESS datasets.  \label{fig:tess5res2}
    }
\end{figure*}

\cite{Southworth_2021} performed a frequency analysis of pulsations in the Sector 6 TESS photometry of VV Ori, recorded in 2018-19. They found 51 significant frequency components between 1.4 and 27.8 d$^{ -1}$, with two dominant pulsation modes ($\nu_1$ = 9.1766 $\pm$ 0.0001 and $\nu_2$ = 9.0324 $\pm$ 0.0002 d$^{ -1}$).  There were also several independent g-mode frequencies below 3 d$^{ -1}$. These components were attributed to $\beta$ Cep oscillations of the primary and SPB variations of the secondary, respectively. 

Here, we have analysed Sector 32 pulsations recorded in 2020, using a different approach to removing the close binary contribution to the LC. Fig.~\ref{fig:tess5res1} 
shows part of the simple aperture photometry (SAP) LC. For clarity, only the first 2.5 days are presented; however, all 27 days' data were analysed. Only the out-of-eclipse observations (shown in red) were used in the frequency analysis, in order to avoid complications to the pre-whitening operation arising from eclipse effects. The proximity effects were removed empirically, by fitting 6th-order polynomials to the out-of-eclipse sections of the phased LCs. The maxima before and after primary eclipses were fitted separately. Residual pulsations after subtracting the polynomial models are shown in Fig.~\ref{fig:tess5res1b}. We used the same methodology to analyse the Sector 6 data, allowing direct comparison with the results from \cite{Southworth_2021}.


Frequency analysis of residual pulsations was performed using {\sc Period04} software \citep{Lenz_Breger_2004}. Iterative pre-whitening identified 60 significant frequencies with amplitude signal-to-noise $>$ 4. The frequencies, amplitudes, and phases are listed in Table~\ref{tab:App1}.  Fig.~\ref{fig:tess5res2} presents the frequency spectra based on the residuals from the LC fittings to the 2018 and 2020 data. General features of the Sectors 6 and 32 amplitude spectra are similar. The two dominant frequencies reported in \cite{Southworth_2021} remain and with similar amplitudes. The frequencies we obtain are: (2018) $9.17680 \pm  0.00007$ and $9.03131 \pm 0.00016$ d$^{-1}$; and (2020) $9.17726 \pm  0.00007$ and $9.03282 \pm 0.00017$ d$^{-1}$. The relative amplitudes of peaks between 9 and 14 d$^{-1}$ are different, however, while the cited frequencies differ by more than their formal uncertainty estimates.  This may be attributed to the different data extraction methods employed, but slight changes in the pulsation properties in the two years between the TESS observation sets cannot be ruled out in view of the low level of the  formal errors. 


\section{Spectrometry}
\label{section:spectrometry}

Spectroscopic data examined in this study include observations made  with the {\sc hercules} spectrograph  \citep{Hearnshaw_2002}, using the 1m McLellan telescope at the University of Canterbury Mt John Observatory (UCMJO) ($\sim$ 43{\degr}59{\arcmin}S, 174{\degr}27{\arcmin}E) in New Zealand.  Around 25 spectral images were obtained, distributed over the period 2010-15. These UCMJO observations provide  somewhat incomplete coverage for the first half  of the full cycle, but allow a consistent result.  The data were collected with a 4k$\times$4k Spectral Instruments (SITe) camera \citep{Skuljan_2004}. Starlight from the telescope was passed to the spectrograph by a 100 $\mu$m effective diameter fibre, corresponding to a theoretical resolution of $\sim$40,000. Wavelength and relative flux calibration was performed using the latest version of the software package {\sc hrsp} \citep{Skuljan_2020} that outputs measurable data in {\sc fits} \citep{Wells_1981} formatted files. Fair weather exposures  were usually for $\sim$500 seconds. 

Some 45 clear orders (80 to 125) of the \'{e}chelle were set up for inspection, using the software package {\sc vspec} (\citealt{Desnoux_2005})\footnote{This is an MS-Windows$^{\rm TM}$ based package that includes essential data-processing functions.}. Useful spectra typically have a signal to noise ratio (S/N) of $\sim$100 in order 85. Given the high  red-sensitivity of the SITe camera, this drops to $\sim$50 by order 125. 
 
Table~\ref{tab:tbl-3}\footnote{Placed in the appendix on page~\pageref{tab:tbl-3} to avoid disrupting the flow of the paper.}
lists individual lines, mainly of the primary, detected in this study. Apart from H$_{\alpha}$  and H$_{\beta}$, the best defined line is probably the primary He I $\lambda$6678 line, where the corresponding weak secondary feature is also seen. A similar situation holds at He I $\lambda$5875; but the other He lines appear
weaker and the secondary is not clearly distinguished at $\lambda$5047. High excitation lines of C, N, O and Si are increasingly seen in the higher orders, though their S/N deteriorates and blending can be  expected as the primary non-hydrogen lines are  $\sim$6-8 \AA\ wide.  
The He II feature at $\lambda$4686, together with the  high-excitation lines, suggest the type may be a bit earlier than the oft-cited B1. 
The interstellar Na II lines reported by \cite{Terrell_2007} are confirmed and separated into two main concentrations separated by about 0.33 {\AA} ($\sim$17 km s$^{-1}$). 

The UCMJO data on VV Ori were combined with a set of 45 high-resolution spectra taken in 2006 and 2007 with the {\sc fies} spectrograph at the Nordic Optical Telescope (NOT) at the Roque de los Muchachos Observatory, La Palma, Canarias, Spain. This facility has been reviewed by \cite{Telting_2014}.  Although some disparities were encountered in combining the {\sc fies} and {\sc hercules} data, the sharp interstellar sodium lines are quite prominent and useful for checking wavelength calibration. The greatest deviations show up in the last two spectra obtained with {\sc hercules}. These images from Dec 2015 are  separated by over a year from the main cluster of observations (Table~\ref{tab:tbl-7}). The {\sc fies} spectra form a more compact dataset, coming from two short observing runs in Nov 2006, and Oct 2007. In the context of possible secular changes in RV curves, it would be desirable to complete phase coverage over as short an interval as possible.
\begin{table}
\begin{center}
\caption{Radial velocity data of the components of VV Ori derived from the He I lines. 
\label{tab:tbl-7}} 
\begin{footnotesize}
\begin{tabular}{llrr} 
\hline

\multicolumn{1}{l}{BJD}  & \multicolumn{1}{l}{Orbital} & 
\multicolumn{1}{r}{RV1} & \multicolumn{1}{r}{RV2}   \\
\multicolumn{1}{l}{2450000+}  & \multicolumn{1}{l}{phase} &
\multicolumn{1}{r}{km s$^{-1}$} & \multicolumn{1}{l}{km s$^{-1}$}\\ 
\hline 
     	5539.9398  &	 	0.4272  & $-26.83$	&		---     \\
  	 	5539.9576  &		0.4391  & $-24.28$	&		---     \\   
  	 	5544.8758  &		0.7502  & 	164.45	&	   $-302.3$  \\
  	 	5544.8917  &		0.7609  &	163.44  &      $-302.3$  \\ 	
            5544.9560  &		0.8040   &	146.60	&	   $-66.1$   \\
 	 	5545.0213  &		0.8482  &	134.39	&	   $-192.6$  \\
 	 	5549.9023  &		0.1342  & $-97.22$	&	 	---      \\
 		5798.2189  &		0.3083  & $-127.32$	&	 	282.3    \\
 	 	5798.2600  &	  	0.3360  & $-120.18$	&		276.7    \\
 		5798.2737  & 		0.3452  & $-120.69$	&	 	280.8    \\
            5875.9380  &		0.6311  &	157.24	&	   $-190.7$  \\
		5875.9391  &	 	0.6318  &	133.34	&	 	$-221.2$ \\
     	5875.9529  &	 	0.6411  &	142.01	&	 	$-226.8$ \\
 		5875.9530  & 		0.6411  &	157.22	&	   $-208.6$  \\	
  		5876.0580  &		0.7119  &	181.71	&	   $-58.2$   \\
 		5876.0584  &	 	0.7121  &	153.74	&	 	$-282.9$ \\
	 	5876.0720  &	 	0.7213  &	181.67	&	   $-260.5$  \\
  		5876.0724  &	 	0.7216  &	159.35	&	 	$-266.1$ \\ 
  
		5880.0972  &	 	0.4312  & $-15.61$	& 		---      \\
 		6255.1256  &	 	0.9114  &	75.7	&	 	---      \\
 		6258.0204  &	 	0.8602  &	109.36	&	 	$-231.9$ \\ 
 		6258.0690  &	 	0.8930  &	101.71	&	 	$-236.0$ \\
  		6666.9080  & 		0.1355  & $-95.60$	&		282.5	 \\ 	 
 		6673.9112  &	 	0.8502  &	106.81	&	 	$-240.0$ \\
 		6675.0142  &	  	0.5928  &	107.71	&	 	$-187.0$ \\
    	6993.9697  &	 	0.3231  & $-194.65$ &	 	202.7    \\
 	 	7350.1629  &	  	0.1229  & $-99.01$	&	 	261.9    \\
 		7356.1382  &	 	0.1456  & $-109.98$	&	 	268.8	 \\ 		 	 	
\hline 
\end{tabular}
\end{footnotesize}
\end{center}
\end{table}

\begin{table}
\begin{center}
\caption{
The orbital parameters for VV~Ori determined by spectral disentangling. 
The orbit is essentially circular, hence only the RV  semi-amplitudes 
$K_{\rm 1}$, and $K_{\rm 2}$ of the components are determined (see Fig~\ref{fig:spd-orbit}). The derived values of the mass ratio, $q$, and masses of the components, $M_{\rm 1}$, and $M_{\rm 2}$ multiplied by $\sin^3\,i$, are given.
\label{tab:rv_fit}} 
\begin{tabular}{lrr}
\hline 
\multicolumn{1}{l}{Parameter}  & \multicolumn{1}{r}{Value}  & 
\multicolumn{1}{r}{Uncertainty}\\
\hline \\
$K_{\rm 1}$  \kms                   &  137.20            & $\pm$ 0.59     \\
$K_{\rm 2}$  \kms                   &  328.58            & $\pm$ 1.56     \\
\hline
$P$   (d)                           &    1.48537742      &  ---            \\
$q$                                 &  0.418             & $\pm$ 0.003     \\
$a \sin i$                          &   13.67            & $\pm$ 0.05      \\
$M_{\rm 1}\,\sin^3\,i$    \Msun     & 10.97              & $\pm$ 0.13      \\
$M_{\rm 2}\,\sin^3\,i$    \Msun     & 4.58               & $\pm$ 0.04      \\
$V_\gamma$ (km s$^{-1}$)            & 26.7               & $\pm$ 2.0       \\
\hline
\end{tabular}
\end{center}
\end{table}


\subsection{Spectral disentangling: determination of the orbital parameters}
\label{section:spectral_disentangling}

Spectral lines in high-mass binary systems, like VV~Ori, are usually broadened by high rotational velocities and often become blended over the course of the orbital cycle. Therefore, direct RV measurements are probably uncertain. In such cases, even more modern cross-correlation function (CCF) techniques become ineffective.

Spectral disentangling \citep{Simon_Sturm_1994, Hadrava_1995} overcomes most of these problems, enabling a simultaneous determination of the orbital parameters, together with a reconstruction of the individual spectra of the components. There is no need for template spectra in this technique, which are usually the main source of uncertainty in RV evaluation by the CCF method.  This can come from mismatches in spectral types \citep{Hensberge_Pavlovski_2007}. Precision is thus gained in the relative RV values, but the mean motion of the system $V_{\gamma}$ has to be determined separately. With disentangling, the components' spectra are effectively separated, which, in turn, permits useful atmospheric diagnostics for either star. This allows a determination of their metallicity values, or detailed abundance signatures. For recent applications of spectral disentangling in complex high-mass binary systems see \citet{Pavlovski_2018,Pavlovski_2023}, where the methodology used in the present work is described in detail.


\begin{figure}
\centering
\includegraphics[width=8cm]{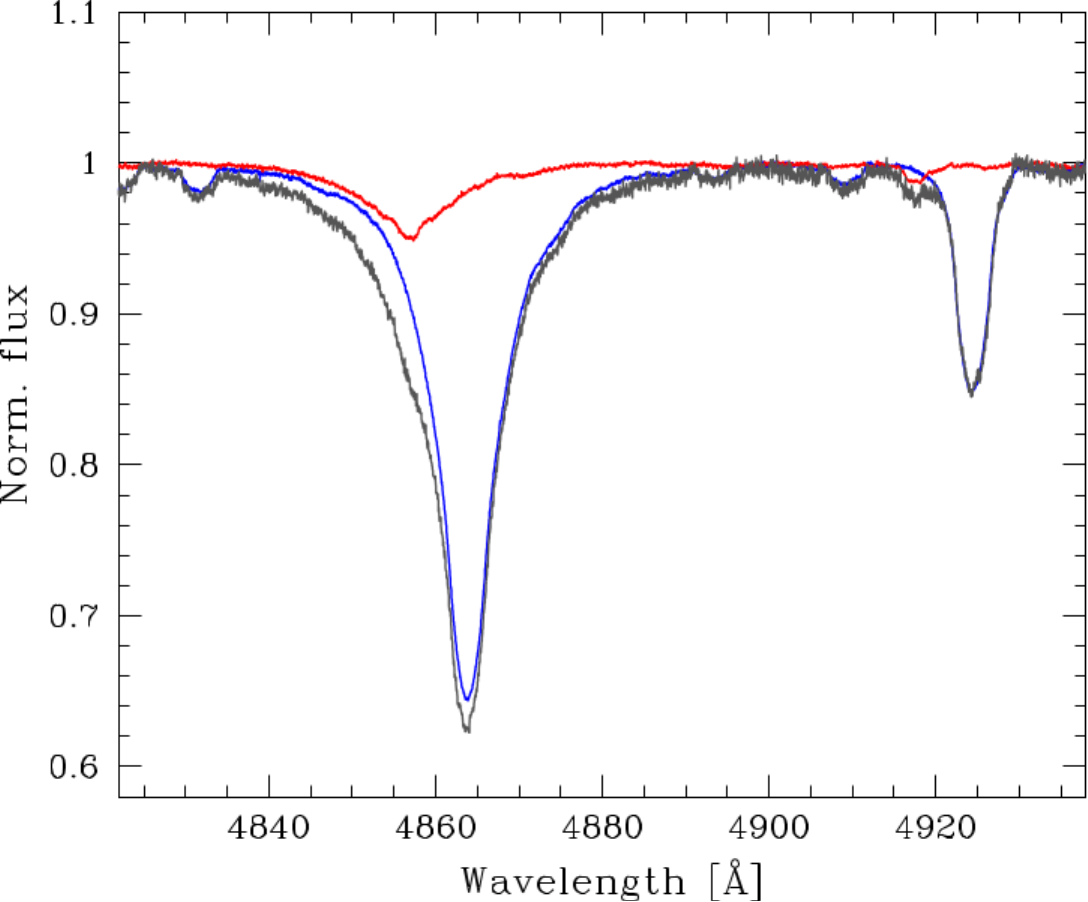}
\caption{Disentangled spectra of the primary component (in solid blue colour) and the secondary component (in solid red colour) superimposed on the observed spectrum of VV~Ori in the quadrature (in solid grey). Whilst the \ion{He}{i} 4920 {\AA} line is clearly resolved, the broad H$\beta$ line shows only an asymmetric profile, due to unresolved components.    
\label{fig:hbeta}
}
\end{figure}

\begin{figure}
\centering
\includegraphics[width=8cm]{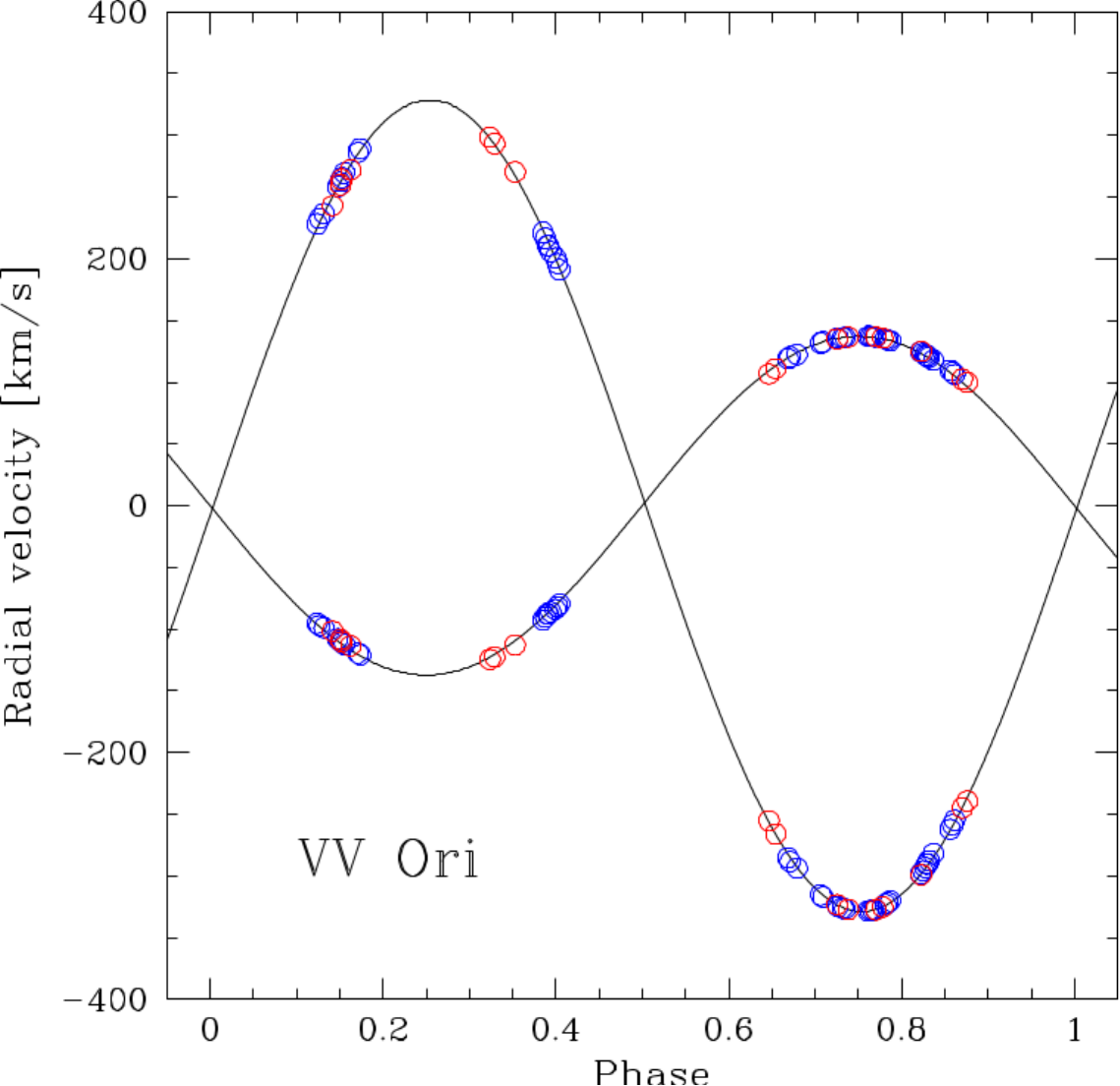}
\caption{The relative radial velocities for the components of VV~Ori representing the orbital solution obtained with spectral disetangling (solid black lines). The observations obtained with the {\sc fies} spectrograph  are shown as open blue circles, while open red ones represent the {\sc hercules} data.   \label{fig:spd-orbit}
}
\end{figure}

Since the {\sc fies}, and {\sc hercules} spectra do not cover all the same spectral range, we concentrated on the region where they overlap, i.e.\ between the H$\gamma$ and H$\beta$ lines. This spectral segment covers about 400~{\AA} and contains various metal lines, the most prominent of which is the \ion{He}{i} 4471~{\AA} line. As seen in Fig.~\ref{fig:hbeta} the helium and metal lines are resolved in the spectra of VV~Ori obtained at quadrature. However, in spite of the large RV amplitude of the secondary component (about 660~\kms), H$\beta$ and the other hydrogen lines reflect this only in a changing asymmetry through the course of the orbital cycle. The two main components are never clearly resolved in the Balmer lines. Although spectral disentangling will resolve the  hydrogen lines of the primary and secondary, because of the severe blending, precision in determination of the orbital parameters from the H lines is certainly smaller than for the resolved helium and metal lines.  

The third body identified through speckle interferometry in the WDS catalogue  is approximately 4~mag fainter in V than the primary. It is therefore probably an early A-type main sequence star \citep{Eker_2018} whose observed spectra would show only relatively weak features apart from the Balmer lines.  Admission of a  third contribution into the modelling at the level of  a few \% of the main component cannot be definitely  confirmed from the residuals in the two component fitting.

The spectral disentangling program specifically referred to in this paper is {\sc FDBinary } \citep{Ilijic_2004}.  This is based on the formulation of disentangling in Fourier space, after \cite{Hadrava_1995}.  Methodological principles concerning disentangling in the wavelength domain are from \cite{Simon_Sturm_1994}. In {\sc FDBinary} Fast Fourier Transforms (FFTs) are used that enable high flexibility in the selection of suitable spectral segments for disentangling whilst maintaining the original spectral resolution. 

The RV curves shown in Figure~\ref{fig:spd-orbit} are based on the line-centre wavelength values determined by the line-fitting process. The observed data-points were  optimally matched by the theoretical model curve (the continuous lines shown in Figure~\ref{fig:spd-orbit}) {derived using the} RV-curve application of {\sc WinFitter}.  {The} corresponding parameters {are} listed in Table~\ref{tab:rv_fit}. The velociy semiamplitudes ($K_{\rm 1}$ and $K_{\rm 2}$) are both significantly larger than those inferred from the masses given by \citet{Terrell_2007}. As a result, the masses we derive are also substantially larger than previously thought.


\begin{table} \centering
\caption{\label{tab:optfit} The main atmospheric parameters for the components
of VV~Ori determined by optimal fitting of their disentangled spectra. Also,
the fractional light contributions of the components (ldf) can be optimised in an
unconstrained mode (see Sect.~\ref{sec:atmos}). } 
\begin{tabular}{lcc} \hline
Parameter              &  Run 1  & Run 2  \\
\hline
   &  Primary &  \\
   \hline
$T_{\rm e}$ [K]       & 26\,010$\pm$320 & 26\,660$\pm$300     \\
$\log g$ [cm\,s$^{-2}$]  &  3.96 $\pm$ 0.02   &   4.083$^{*}$      \\
$v\sin i$  [\kms] &  138.4$\pm$2.3  &  151.4$\pm$2.5  \\
ldf  &   0.908$\pm$0.007  &   0.901$\pm$0.007 \\    \hline 
    &  Secondary  &  \\
\hline
$T_{\rm e}$ [K]       & 15\,780$\pm$420 & 
  16\,250$\pm$410    \\
$\log g$ [cm\,s$^{-2}$]  &  4.29$\pm$0.03          &   4.320$^{*}$      \\
$v\sin i$ [\kms] &  89.3$\pm$3.1  &  98.1$\pm$3.5  \\
ldf  &   0.096$\pm$0.008  &   0.095$\pm$0.008 \\    
\hline
\end{tabular}
\end{table}


\subsection{Atmospheric parameters}
\label{sec:atmos}

Once individual components' spectra have been  separated they can be used for detailed spectroscopic analysis as single star spectra. The disentangled spectra still refer to the common continuum of the binary,  hence they are diluted by the fractional light contribution of either component to the total light of the system.    Generally, there are two options at this point: (i) disentangled spectra might be first re-normalized to their own continua, using the light ratio determined by other means, i.e.\ LC  analysis \citep{Pavlovski_Hensberge_2010}, or (ii) the analysis could be performed directly on the disentangled spectra, since these spectra contain information on the light ratio  \citep{Tamajo_2011}. It has been shown that uncertainties in the determination of the light ratio from disentangled spectra are  comparable to the uncertainties achieved in photometric analysis, and are typically on the order of 1 percent \citep{Pavlovski_2009, Pavlovski_2018, Pavlovski_2022, Pavlovski_2023}. In the present work,  we decided on the second option, in which the light ratio pertaining to the spectral segment studied would be determined simultaneously with the atmospheric parameters. 

The optimal fitting of our disentangled spectra was performed by an extensive search through a pre-calculated grid of theoretical spectra. As a fitting merit indicator we used the sum of squared residuals between the disentangled spectrum and the selected synthetic spectrum using the {\sc starfit} \citep{Kolbas_2014} code. The search  procedure is performed by a genetic algorithm, that is based on the {\sc pikaia} subroutine of \citet{Charbonneau_1995}. The uncertainties are calculated using a Markov Chain Monte Carlo (MCMC) procedure  \citep{Ivezic_2014}.

\begin{figure*}
\centering
\includegraphics[width=8cm]{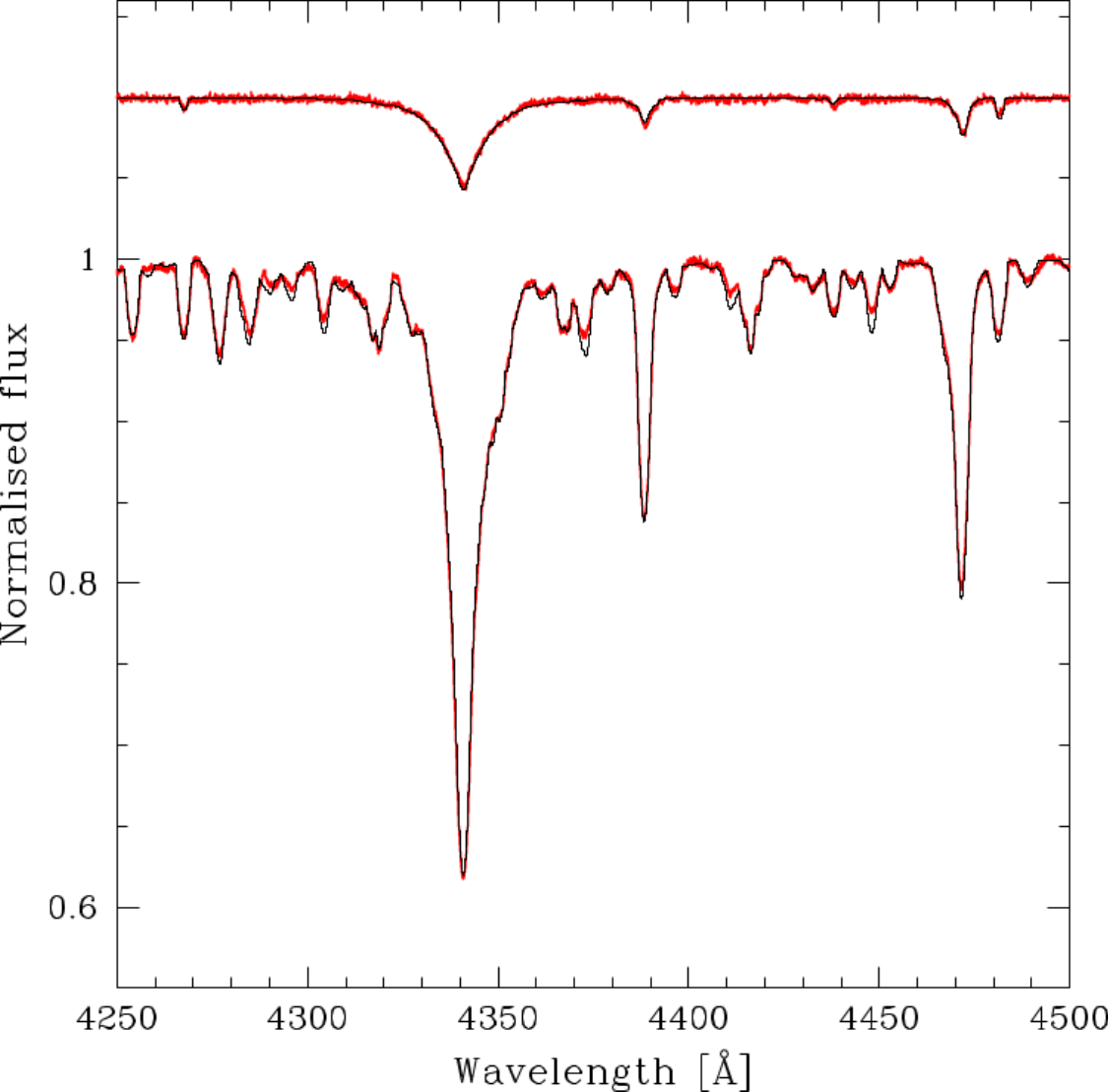}  \includegraphics[width=8.12cm]{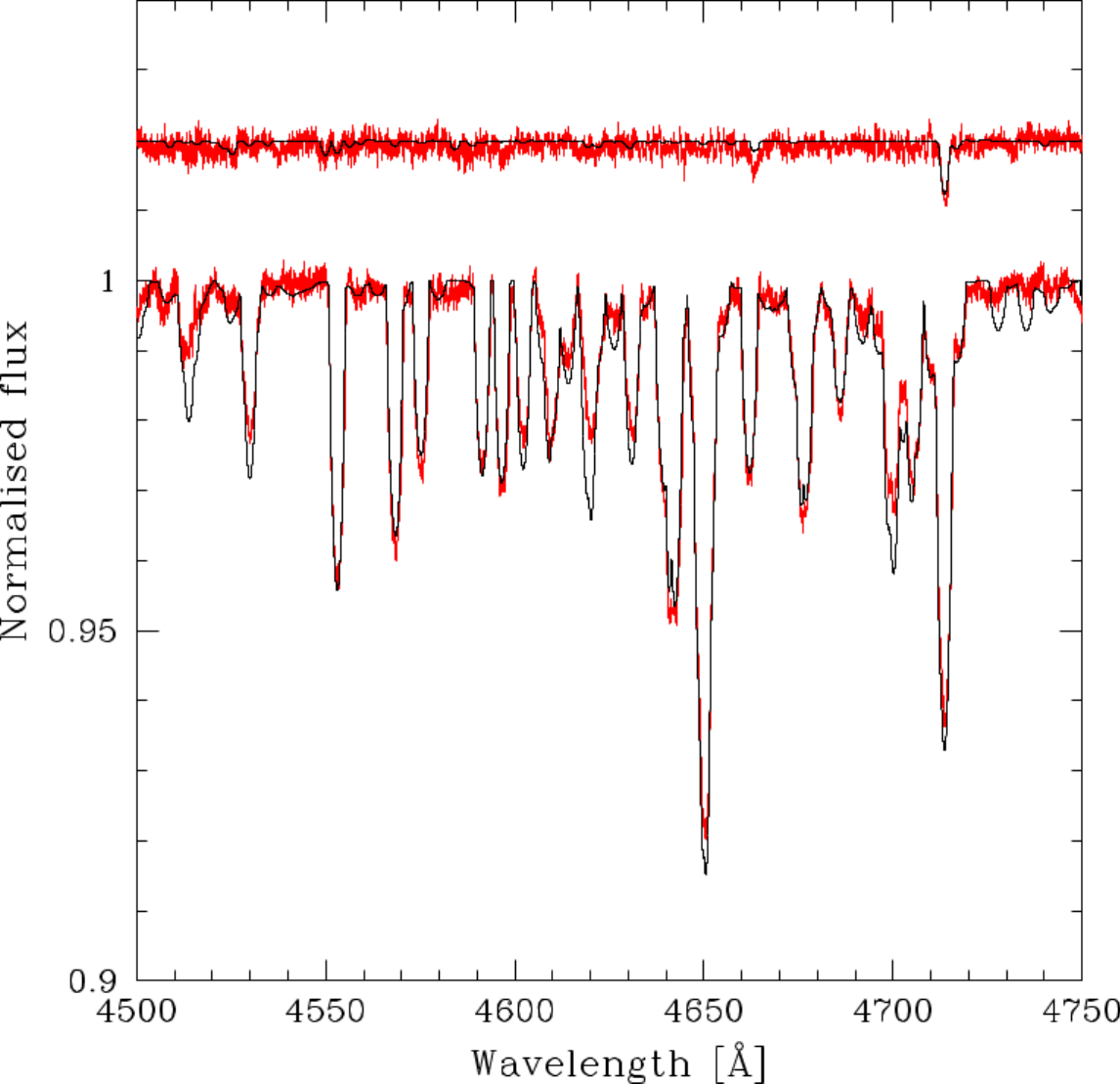} 
\caption{Optimal fitting of disentangled spectra for the primary component (lower) and secondary component (upper). Disentangled spectra are in red, optimal fits in black. The region of the spectra centred on the H$\gamma$ line (left panel) and various metal lines in the spectral range from $\lambda$ 4500 - 4750 {\AA} (right panel) are shown.    
\label{fig:fit_spec}
}
\end{figure*}


Since, in a close binary system, the sum of the fractional light contributions of both components should be unity, we would normally constrain the optimal fitting of the disentangled spectra with that condition. 
However, as discussed above, there are indications of a faint companion to the close pair in VV~Ori. We therefore performed the optimal fitting in an unconstrained mode to allow for a  light contribution of such a third star. The result of the modelling, with all atmospheric parameters and fractional light contributions left free, are given in Table~\ref{tab:optfit}. Model fits can be seen in Fig.~\ref{fig:fit_spec}.

The sum of the light dilution factors (ldf), i.e.\ the fractional light contributions of the components to the total light of the system, determined by our optimal fitting of disentangled components' spectra then became slightly larger than unity, thus denying any significant light contribution from a  third component. This finding does not support the results of the LC analysis in the present work. In the LC analysis of BVR photometry of VV~Ori (Table~\ref{tab:congarinni_fit}), and the TESS observations (Table~\ref{tab:tess_fit}) a small third light appeared, in agreement with the astrometric findings of \cite{Horch_2017}.    
The light ratio of the main components we derived from the foregoing spectroscopic analysis, however, $lr_{\rm sp}$ = $l_{\rm 2}/l_{\rm 1}$ = $0.106 \pm 0.009$ is in reasonably good agreement with Terrell et al.'s photometric light ratio in the B and b passbands, i.e.\ $lr_{\rm ph} = 0.093 \pm 0.001$, and $0.098 \pm 0.001$, respectively.

The eclipsing, double-lined spectroscopic binary nature of VV Ori's close pair allows determination of their masses and radii with  high precision. Since one of the main obstacles in determination of the atmospheric parameters from hydrogen line profiles is degeneracy between the $T_{\rm e}$ and surface gravity, a determination of the latter from the system's dynamics could be used to lift this degeneracy.  In the second run of the optimal fitting of the disentangled components' spectra, the surface  gravity values for both components were fixed to those determined from combined spectroscopic and LC analyses. The results of Run 2 with $\log g$'s thus fixed, as indicated by an asterisk, are given in Table~\ref{tab:optfit}. Breaking the degeneracy in the $T_{\rm e}$ and $\log g$ parameters has significant effects on the $T_{\rm e}$ values.  These turn out to be about 650, and 250 K higher than the first run estimates for the primary and secondary components, respectively.   

The primary's $T_{\rm e}$ determined by \citet{Terrell_2007} was $T_{\rm e,1} = 26\,200$ K.  This is in fair agreement with our determinations in both runs. 
This finding is encouraging, since \citet{Terrell_2007} based the value of the primary's $T_{\rm e}$ on their de-reddened B-V, using an estimate of the interstellar reddening from Na~I lines. Older estimates of the $T_{\rm e}$ for the primary component (Table~\ref{tab:history}) were based on \citet{Eaton_1975}, who found $T_{\rm e,1} = 25\,400\pm1500$ K from
modelling the spectral energy distribution in the UV. 


\subsection{Rotational velocities}
\label{sec:rotational}

If the resolution is sufficiently high, spectral line profiles can be modelled with a parameter set that determines the source's rotation rate  and scale of turbulence in the surrounding plasma, as well as  the wavelength of the centre of light. Such modelling has been carried out in numerous previous studies \citep{Shajin_1929, Huang_1954, Slettebak_1985, Butland_2019}.
 The profiles of the He I lines in our high-dispersion spectral images, particularly the $\lambda 6678$ feature, are well suited to this purpose. 
 
\begin{table}
\begin{center}
\caption{Line-modelling parameters for VV Ori averaged from out-of-eclipse observations (see Section~\ref{sec:rotational} for explanation of the parameters).   
\label{tbl-6}} 
\begin{tabular}{lcccc}
\hline  
\multicolumn{1}{c}{Parameter}  & \multicolumn{1}{c}{primary} & \multicolumn{1}{c}{$\pm$} & \multicolumn{1}{c}{secondary}
& \multicolumn{1}{c}{$\pm$} \\
\hline 
$U$                             & 0.954         & 0.001         & 0.955         & 0.001 \\
$I_{0}$                         & --0.101       & 0.004         & $-0.018$      & 0.002\\
$v_{\rm rot}$ km s $^{-1}$      & 147.1         & 5.7           & 86.1          & 10.3\\
$s$ km s$^{-1}$                 & 5.6           & 2.3           &34.3           & 1.6 \\
$\Delta f$                      & 0.007         &               & 0.007         &\\
$\chi^2/\nu $                   & 1.0           &               &0.93           &  \\
\hline
\end{tabular}
\end{center}
\end{table} 

\begin{figure}
\includegraphics[width=\linewidth]{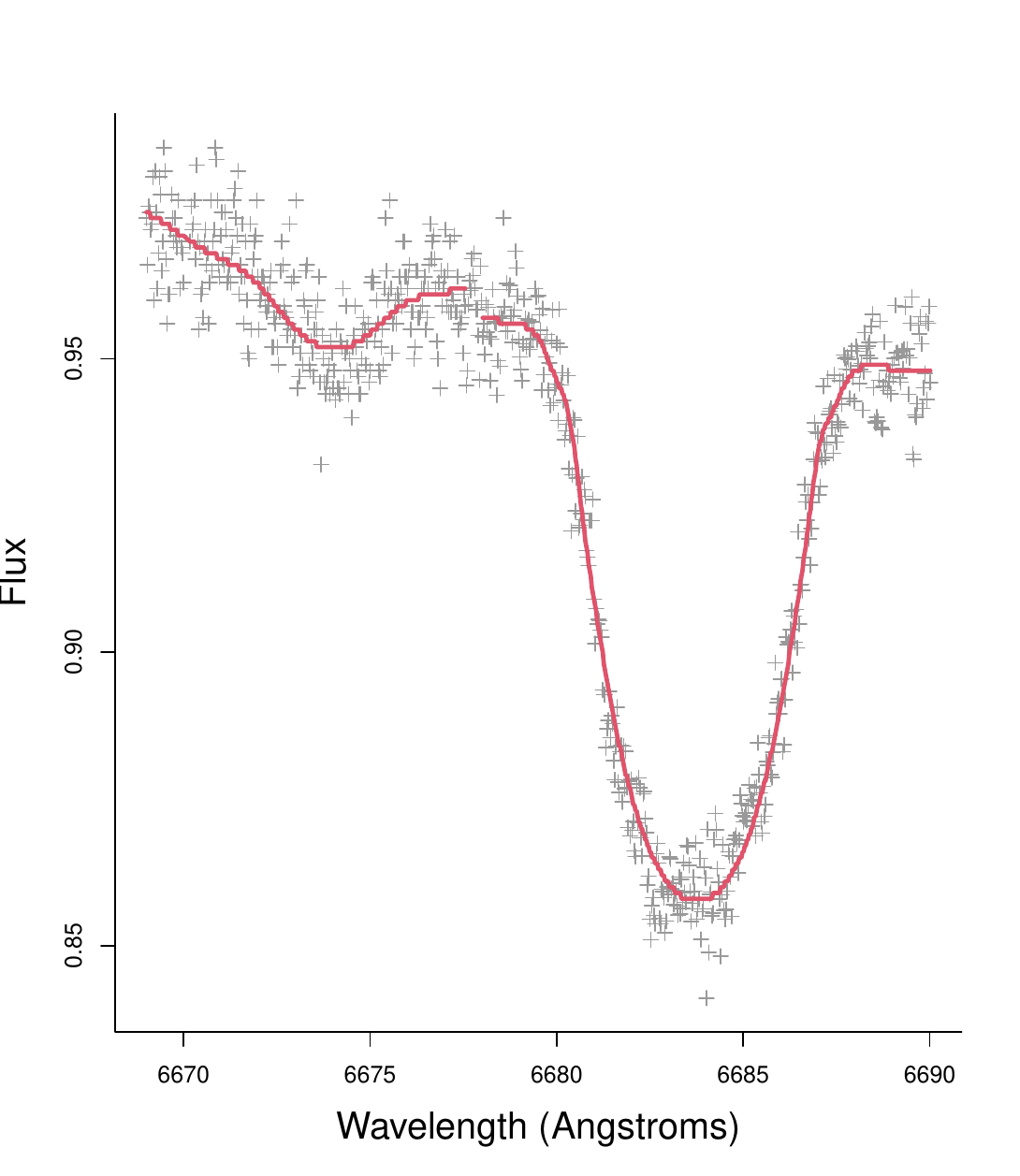}\\
\caption{Profile fitting to the He I $\lambda$6678 lines: 
The lines have been fitted separately and do not quite return to the same sloping continuum level.
\label{fig:profile_fitting}
}
\end{figure} 

\begin{table} 
\centering 
\caption{\label{tab:abundances}
        Rotational velocities: observed and calculated} 
        \begin{tabular}{lcc} \hline
                            &   {\rm Primary}       &  {\rm Secondary} \\
        \hline
        $v \sin i$          &  $138.4 \pm 2.3$      & $89.3 \pm 3.1$ \\
        $v_{\rm sync} $     &  $176.8 \pm 0.8$      & $86.4 \pm 1.7$ \\
    \hline 
    \end{tabular}
\end{table}
 
The entries in Table~\ref{tbl-6} follow a similar arrangement to the other tables of model-fitting in this paper. $U$ refers to the continuum reference level at the line's central wavelength corresponding to the example shown in Fig.~\ref{fig:profile_fitting}. The local continuum is assumed to be a straight line, but it may have a definite slope in the raw data
that is empirically dealt with in the fitting procedure.
The quantities $I_0$ correspond to the central depths  relative to the local continuum (average of the six profiles examined). 
The equatorial rotational velocity of the primary $v_{\rm rot, 1}$  is calculated with an inclination $i = 79.0^{\circ}$ (Table~\ref{tab:tess_fit}) to correct for the projection. The parameter $s$ relates to  the scale of gaussian broadening that is convolved with that from the rotation.  It may be interpreted as a measure of the turbulence of the source plasma, or perhaps some other near-symmetric broadening effect. The conspicuous, and consistently obtained high value of this parameter for the secondary (compared to the primary)  is noteworthy. $\Delta f$ indicates the standard deviation of 
raw measures in the data sample.
 
 
For synchronized rotation with aligned axes we would expect  the projected equatorial speed of the primary, using Table~\ref{tab:abundances}, to be $\sim 0.368 \times 466/ \sin i$ km s$^{-1}$, i.e. $\sim $ 175 km s$^{-1}$, but from Table~\ref{tbl-6} it is seen that the observed speed  of 147 km s$^{-1}$ is significantly  less than that.
The ratio of the two rotational speed estimates ($\sim$0.59) is also significantly different to that of the ratio of radii following from Table~\ref{tab:tess_fit} (0.492). In fact, if we divided the secondary's rotational velocity by that ratio we would obtain the synchronous rotation value 175 km s$^{-1}$ for the primary, i.e.\ the mean rotational speed of the secondary is close to synchronism, unlike the primary.

Following the discussion  of \cite{Southworth_2021}, this can be interpreted as a non-alignment of the primary's rotation and orbit axes, that would correspond to an obliquity angle ($ \epsilon $) at zero precession angle ($ \psi $) of around 33$^\circ$. Precession of this spin axis could cause the apparent variation of  inclination reported by \cite{Southworth_2021}.
   

\subsection{Elemental abundances for the primary component}
\label{section:elemental_abundances}

As the primary component contributes most of the light from this binary, it has more weight in the spectral disentangling. The S/N of the primary's disentangled spectrum is thus relatively high. This is an important point for detailed abundance analysis. Our adopted procedure is fully described in  \citet{Pavlovski_2018, Pavlovski_2023}. A grid of NLTE (non-local thermodynamic equilibrium) synthetic spectra is calculated using the programs {\sc detail} and {\sc surface} with model atmospheres first produced by {\sc atlas9} using the LTE (local thermodynamic equilibrium) prescription. Theoretical spectra are broadened according to a given instrumental profile and projected rotational velocity. Abundances are then determined from line profile fittings of the normalised disentangled spectrum of the primary to the adjustable theoretical one. 

The following species were considered: C, N, O, Mg, Si, and Al. The microturbulent velocity, $\xi = 2\pm1$ {\kms}, is determined from the oxygen lines, which appear to be the most numerous in the primary's spectrum.  The result of the abundance analysis is given in Table~\ref{tab:abundances_2}. Determined abundances are in excellent agreement with previous analyses of high-mass stars in detached binary systems (see \citealt{Pavlovski_2018, Pavlovski_2023}).  There has been no previous abundance analysis specifically for VV~Ori, but stars in the Ori I association  have been extensively studied through recent decades (cf.~\citealt{Simon-Diaz_2010, Nieva_2011}, and references therein). Clearly, the elemental abundances, and nitrogen-to-oxygen (N/O) and nitrogen-to-carbon (N/C) abundance ratios for the primary component in VV~Ori is the same as for stars in the Ori~I association, within the given  uncertainties. 

\begin{table} \centering
\caption{\label{tab:abundances_2} 
Photospheric elemental abundances for the primary component in VV~Ori. For comparison, abundances determined for B-type stars in the Ori I association in \citet{Nieva_2011} (abbreviated NSD) are also presented. The elemental abundances for species X is given relative to the hydrogen abundance, $\epsilon(\rm{X}) = \log({\rm X}/{\rm H}) + 12$. The N/O and N/C abundance ratios are also given, as a sensitive probe to mixing processes in the stellar interiors.}
\begin{tabular}{lcc} \hline
Element  & This work      &  NSD    \\
        &   [dex]         &   [dex]  \\
\hline
C      & $8.28 \pm 0.06$    &  $8.35 \pm 0.03$      \\
N      & $7.77 \pm 0.07$    &  $7.82 \pm 0.07$      \\
O      & $8.76 \pm 0.07$    &  $8.77 \pm 0.03$      \\
Mg     & $7.64 \pm 0.06$    &  $7.57 \pm 0.06$      \\
Si     & $7.51 \pm 0.08$    &  $7.50 \pm 0.06$      \\
Al     & $6.41 \pm 0.08$    &   -                 \\
N/O    & $-0.99\pm 0.10$   &  $-0.95 \pm 0.08$    \\
N/C    & $-0.51\pm 0.09$   &  $-0.53 \pm 0.08$     \\
\hline \\
\end{tabular}
\end{table}

\section{Absolute Parameters}
\label{sec:absolute_parameters}

\begin{table}
\caption{Adopted absolute parameters of VV Ori,
derived from the combined photometric and spectroscopic analyses discussed in Sections 2 and 3.
 The units labelled with an `N' are given in terms of the nominal solar quantities 
defined in IAU 2015 Resolution B3 \protect\citep{Prsa_2016}. 
\label{tab:absolute_parameters}}
\begin{center}
\begin{tabular}{llrr}
\hline
\multicolumn{2}{l}{Parameters} & \multicolumn{1}{r}{Value} & 
\multicolumn{1}{r}{Uncertainty} \\ 
\hline
Masses (\Msunnom) & $M_{1}$   & 11.56              &   0.14           \\ 
& $M_{2}$                     & 4.81               &   0.06           \\
& $M_{3}$                     & 2.0                &   0.3            \\
Radii (\Rsunnom) & $R_1$      &  5.11              &   0.03          \\ 
& $R_2$                       &  2.51              &   0.02          \\
& $R_3$                       &  1.8               &   0.10           \\
Semi-major axis (\Rsunnom) & $a$ & 13.91             &   0.05           \\
Temperatures (K) & $T_{e1}$   & 26660              &   300            \\ 
& $T_{e2}$                 & 16250              &   420            \\
& $T_{e3}$                & 10000              &   1000           \\ 
Luminosities $\log(L/\Lsunnom)$ & $\log L_1$        &  4.07    &   0.02  \\
& $\log L_2$                  &  2.60              &   0.06           \\
& $\log L_3$                  &  1.5               &   0.2            \\
Absolute magnitudes & M$_{{\rm bol}_1}$ &  $-5.44$     &   0.05           \\
& M$_{{\rm bol}_2}$           &  $-1.75 $          &   0.14           \\ 
& M$_{{\rm bol}_3}$           &  1.1               &   0.40           \\ 
Reddening (mags) &  $E(B-V)$         & 0.08   &   0.03              \\
Gravities  ($\log$[cgs])& $\log{ g_{1}}$   &  4.08  &   0.05     \\
& $\log{ g_{2}}$              &  4.32              &   0.06           \\ 
& $\log{ g_{3}}$              &  4.25              &   0.10            \\
Distance (pc) & $\rho$        &  396               &   7            \\
\hline
\end{tabular}
\end{center}
\end{table}

It is well known that the actual sizes of the component stars in an eclipsing binary system (radii $R_1$, $R_2$) can be determined by combining the results of LC and RV curve parametrization  --- the `eclipse' or `Russell's' method.  The absolute size of the orbit comes from dividing the RV parameter $a \sin i$ by $\sin i$, using the inclination $i$ from  the LC modelling. The sought radii are simply the product of $a$ and the fractional radii $r_{1,2}$. Surface gravities ($g_{1,2}$), are scaled from the solar value ($\log g_{\odot} = 4.437$) using the radii and masses $M_{1,2}$ similarly derived from the combined RV and LC parameters, or using Kepler's third law, given the period $P$ and separation $a$ of the two stars.

We thus determined the physical properties of VV Ori from the photometric results of Section~\ref{section:photometry}  and the spectrometry of {Section~\ref{section:spectrometry}. For this we have used the {\sc jktabsdim} code \citep{Southworth_2005}, modified to apply the IAU system of nominal solar values \citep{Prsa_2016}, together with the NIST 2018 values for the Newtonian gravitational and the Stefan-Boltzmann constants.  Error-bars were derived from the perturbation analysis referred to in Section~\ref{section:light_curve_analysis}.  The results are given in Table\,\ref{tab:absolute_parameters}.

We calculated a distance to the system using optical $UBVR$ magnitudes from \citet{Ducati_2001}, near-IR $JHK_s$ magnitudes from 2MASS \citep{Cutri_2003}  converted to the Johnson system using the transformations from \cite{Carpenter_2001}, and bolometric corrections from \cite{Girardi_2002}. The interstellar reddening was determined by requiring the optical and near-IR distances to match. We found a distance of $396 \pm 7$ pc.  This  is significantly shorter than the distance of $441 \pm 22$ pc from the \textit{Gaia} DR3 parallax  \citep{Gaia16, Gaia18, Gaia21, Gaia23}. Possible explanations for this are that the Gaia parallax was affected by the brightness of the system and/or the presence of the nearby third body. Evidence in  support of this comes from the renormalised unit weight error (RUWE) of 1.347 being close to the upper limit of 1.4, beyond which the Gaia parallax is considered unreliable\footnote{\texttt{https://gea.esac.esa.int/archive/documentation/GDR2/ Gaia\_archive/chap\_datamodel/sec\_dm\_main\_tables/
ssec\_dm\_ruwe.html}}.

The photometric parallax ($\pi$) can be derived from the formula (\citeauthor{Budding_2007}, \citeyear{Budding_2007}, Eqn~3.42)
\begin{equation}
\log  \pi = 7.454 - \log R - 0.2 {\rm V} - 2F^{\prime}_V  \,\,\,  ,
\label{eq:parallax}
\end{equation}
where $F^{\prime}_V $ is directly proportional to the logarithm of a star's mean surface flux \citep{Barnes_1976}, and is specified by $F^{\prime}_V = \log T_e  + 0.1 BC$, where $BC$ is the bolometric correction. Applying Eqn~\ref{eq:parallax} directly to the  stars in VV Ori, with the V magnitudes from Section~\ref{section:photometry},  the $R$ and $T_e$ values from Table~\ref{tab:absolute_parameters}, and the $BC$ values from \cite{Budding_2007} Table~3.1,  we obtain $\log \pi_1 = -2.64$, and $\log\pi_2 = -2.66$.  The third star would also produce a  comparable value $\log \pi_3 \approx -2.7$,  if we put its $T_{\rm e}$ at 10000 K, 
but data on that star are still very approximate. These parallaxes are in close agreement with the value cited above from the Gaia DR3, but the measured B -- V colour excess argues against adopting the measured V as unaffected by interstellar absorption.  Using the relation of \cite{Cardelli_1989} with the  reddening $E = 0.08$, i.e.\ $A_V = 0.26$, the mean distance turns out to be $\rho = 396 \pm 7$ pc, in good agreement with the foregoing estimate from the V -- I colour.


Evaluation of the absolute luminosities ($L_{1,2}$) and bolometric magnitudes ($M_{{\rm bol} 1,2}$) of the component stars requires the $T_{\rm e}$ values to be known. These were enumerated from the spectral disentangling results given in Table~\ref{tab:optfit}, and checked with the colours derived from Table~\ref{tab:congarinni}. 
In these calculations, we adopted the solar calibration values as: effective temperature $T_e$ = 5780 K, $M_{\rm bol}$ = 4.75
from the IAU-adopted solar constants. The adopted  absolute parameters for VV Ori  are listed, with their uncertainties, in Table~\ref{tab:absolute_parameters}. 

The parameters of the third star are included in Table~\ref{tab:absolute_parameters} for completeness, assuming that it is  coeval with the close binary and about 4.5 V mag fainter (Section~\ref{section:introduction}).  The third star's properties should still be regarded as quite imprecise compared with those of the main components.  Better knowledge of this wide system (VV Ori AB) can be expected in future high accuracy survey work.


\section{Discussion}
\label{section:discussion}
 
Part of the rationale for this study was the ongoing programme of precise quantification of stellar properties, using modern data and analysis techniques. We have shown that independent use of two different LC analysis procedures on recent datasets from the TESS programme resulted in very similar values for the  main geometric parameters.  
Combining these parameters with the RV analysis on high-dispersion {\sc fies} and {\sc hercules} spectrograms, then recovers  absolute parameter sets that are within reasonable error estimates of each other. This bolsters confidence in the employed analytical methods, and provides reliable evidence to check against theory. Fig.~\ref{fig:isochrones} shows such a comparison. The {\sc MESA} Isochrones and Stellar Tracks {\sc MIST} facility \citep{Dotter_2016, Choi_2016} based on the {\sc MESA} models \citep{Paxton_2011} was used to plot evolutionary tracks and isochrones for comparison with our derived parameters. The masses and luminosities are in close agreement with corresponding models at a $\log$ age (in yr)  of $\sim$6.9.} 

Our results confirm that VV Ori is a young binary system and shares properties with other members of the Orion Ib OB star association, with an age between about 6 and 10 Myr. We confirm a photometric distance of around 400 pc: closer to the value ($\sim$360 pc) of \cite{Brown_1994} for the Ib subgroup, but lower than the distance ($\sim$500 pc) of \cite{Warren_1978}.  The Gaia DR3 value ($\sim$440 pc) was considered relatively imprecise, perhaps due to calibration difficulties for this bright star in a crowded field containing nebulosity. Post-Gaia  population studies \citep{Zari_2019} have identified substructures within the Orion Ib Association (see also \citeauthor{Warren_1977}, \citeyear{Warren_1977}; \citeyear{Warren_1978}). Interestingly, the galactic co-ordinates of VV Ori would place it in the B7, or, marginally, the E subgroup, which has an estimated mean age of close to 11 Myr.

\begin{figure}
\includegraphics[width=8.2cm,angle=0]{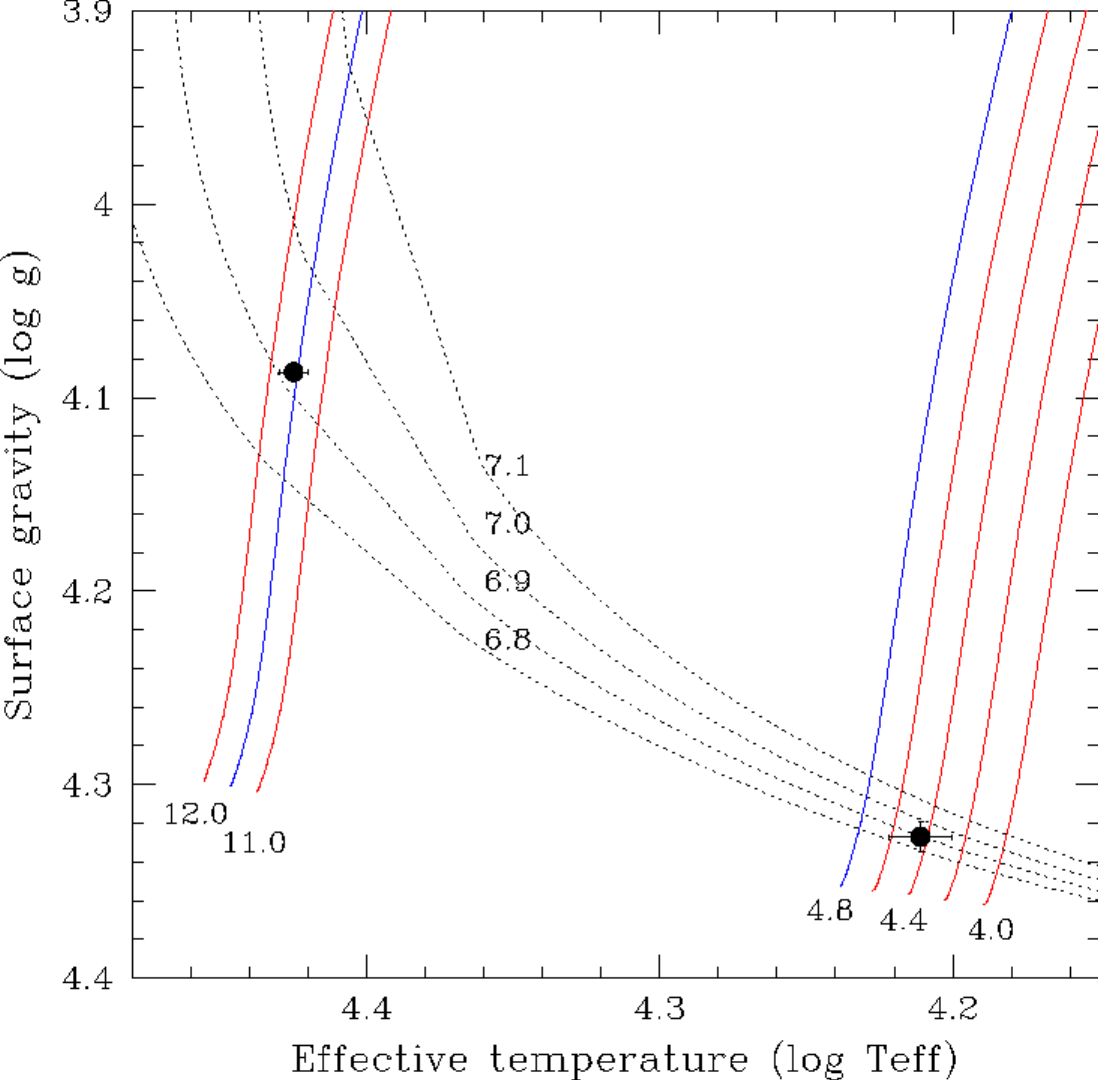}\\
\caption{Derived $\log T_e$  and $\log g$ parameters (Table~12) with their uncertainties for the close pair in VV Ori  (filled black circles) are compared with the MIST evolutionary tracks and isochrones \citep{Dotter_2016, Choi_2016}. The near-vertical evolutionary tracks are labelled according to masses in M$_\odot$. Blue curves are close to the adopted values. Isochrones (dotted) are given in logarithm of time in yr. 
\label{fig:isochrones}
}
\end{figure} 

In Section~\ref{sec:absolute_parameters} we have given the $\log$ age value as $\approx$ 6.9, supporting the idea that VV Ori is slightly older than the bulk of the $\epsilon$ Ori association (see Fig.~\ref{fig:Orion1bcm}). The age of 8 My is, however, young enough to fall within \citeauthor{Zahn_1977}'s (\citeyear{Zahn_1977}) synchronization timescale for massive stars with radiative envelopes and initial separations of around 20 R$_{\odot}$.

In this context, it may be possible to form an independent assessment of the age by referring to the estimated synchronization time-scales. The observed $ v \sin{i}$, and  synchronized velocities are given in Table~\ref{tab:abundances}. The observed  rotational velocity of the primary appears low by $\sim$40 km sec$^{-1}$. The secondary's rotational velocity is, on the  other hand, in good agreement with a synchronized state. The low width of the primary's line core implies that any additional broadening effect, such as macro-turbulence, would not improve agreement between the observed primary rotation and the value corresponding  to synchronization.

This low value of the apparent rotation is, however, in keeping with the possibility of a displaced spin axis raised in Section~\ref{section:fitting_function}. Given the report of changes in the apparent inclination \citep{Southworth_2021} and keeping in mind the interaction between the heat-transfer driven $\beta$ Cep pulsations and tides \citep{Townsend_2018, Pedersen_2022}, the phenomenon of precession can resolve the various observed oddities of the system.  The dynamics of this situation then invites critical attention from relevant theory.

Here we may note that a major source of uncertainty in modelling stellar structure and pulsation properties of upper main sequence stars is the distribution and transport of internal angular momentum. Asteroseismology allows for the determination of the interior rotation of stars, provided non-radial oscillations are present at a suitable level. Normal modes of oscillation, calculated for non-rotating, single stars, become split by rotation and tidal distortions.

Four main processes have been taken to contribute to angular momentum transport within stellar radiative regions: meridional circulation, turbulence driven by instabilities, magnetism, and internal waves \citep{Goldreich_1989, Zahn_2013, Mathis_2013}.  During core hydrogen and helium burning phases, single stars are expected to rotate nearly uniformly. However, in a binary system, even during early phases of evolution, tidal interactions may give rise to non-uniform rotation. In particular, differential rotation in radiative envelopes can induce a diversity of hydrodynamical and magnetohydrodynamical instabilities, that will, in turn, transport and redistribute angular momentum. The secondary component of VV Ori appears to be in near-synchronous rotation, so we expect that differential rotation in that star to be relatively unimportant in comparison with that of the asynchronous primary.

Analysis of the low-order oscillation modes observed in $\beta$ Cep stars should be able to inform us about the star's internal structure.  Thus, convective overshoot at the convective core/radiative interface is a major source of turbulence. Mixing length theory contains a free parameter which determines the extent of overshoot in terms of the local pressure scale height $H_p$. This overshoot region is probed by comparing observed oscillation frequencies with those of models for a range of prescriptions of the overshoot parameter.  So, as well as determining the size and location of the convective core, the frequency spectrum reflects conditions around it that bear on the stellar modelling and angular momentum regime.  

A potentially important source of angular momentum transport is internal gravity  waves (IGW) propagating in the stably-stratified envelope. The generation and damping of these waves depend sensitively on their frequency and length scales. They may be present either at the core-radiative interface, or in thin convection zones near the surface, where local opacity increases in accordance with the $\kappa$-mechanism, or else by tidal forcing \citep{Zahn_1975}. For the latter, tidal forces generate a disturbance at the interface that propagates away from the core. Such waves are large scale $l =2$, $m = 1$ or 2 and have frequency equal to the forcing frequency $\sigma$ (in the co-rotating frame). In a rotating star, $\sigma = k\Omega_{{\rm orb}} - m \Omega$, where $\Omega_{{\rm orb}}$  is the orbital frequency and $\Omega$ that of the rotation; while $k > 0$ and $m$  $(|m| \leq l)$ are integers. Dissipation of these waves occurs by non-adiabatic thermal damping effects near the surface of the star. This is different for prograde $(m > 0)$ and retrograde $(m < 0)$ waves, thus giving rise to a net transport of angular momentum. In this way, IGWs couple convective core and radiative envelope regions, and from asteroseismic data it appears that this coupling is strong.

However, the exact quantification of  such conditions from seismic analysis appears still unsettled regarding modelling or observational data selections, with implications on the confidence of interpretation \citep{Salmon_2022}.  In particular, models that include rotation and/or tidal distortions are not yet directly available.

One source of uncertainty is the extent of chemical mixing which, in young intermediate to high-mass stars arises from instabilities in the radiative layers and, based on interpretation of asteroseismic data, is orders of magnitude lower than expected. Chemical gradients at the core-envelope interface should develop as the star evolves and affect the amplitudes of low-frequency $g$-mode waves. 

With regard to magnetic fields transporting angular momentum, these  may be thought to be largely confined to the convective core \citep{MacGregor_2003}, though other convective zones near the surface would generate fields that could transmit angular momentum along connecting field lines.

\cite{Townsend_2018} have investigated the transport of angular momentum by heat-driven non-radial $g$-modes, producing  a table of models covering most of the mass range of young B-type stars.  From their Fig~7, it can be seen that both VV Ori's primary and secondary (Table~\ref{tab:absolute_parameters}) are on the edge of the region in the HR diagram, in which their torque instability operates. While this may suggest that the proposed mechanism is not viable, the model does not include the effects of tides or rotation. Moreover, \citeauthor{Townsend_2018} assumed aligned spin and orbital axes, whereas with the putative precession in VV Ori this would not be the case. One of the effects of non-alignment would be that the  $m = \pm1$ modes would also contribute to angular momentum transport in the radiative region, in addition to the $m = \pm 2$ modes (for the dominant $l = 2$ tide). 

Clearly, the effects of tides and rotation on the instabilities that give rise to these waves need to be investigated before any definitive predictions can be made. Meanwhile, the juxtaposition of the properties of the stars  in VV Ori, as determined from the classical methods reported in this paper, against  asteroseismological inferences should have very interesting consequences.


\section{Acknowledgements}
 
Generous allocations of time on the 1m McLennnan Telescope and {\sc hercules} spectrograph at the Mt John University Observatory in support of the Southern Binaries Programme have been made available through its TAC and supported by its  Director, Dr.\ K.\ Pollard and previous Director, Prof.\ J.\ B.\ Hearnshaw. Useful help at the telescope was provided by the MJUO management (N.\ Frost and previously A.\ Gilmore \& P.\ Kilmartin). Considerable assistance with the use and development of the {\sc hrsp} software was given by its author Dr.\ J.\ Skuljan, and very helpful work with initial data reduction was carried out by R.\ J.\ Butland.

{ This work was based on observations made with the Nordic Optical Telescope, owned in collaboration by the University of Turku and Aarhus University, and operated jointly by Aarhus University, the University of Turku and the University of Oslo, representing Denmark, Finland and Norway, the University of Iceland and Stockholm University at the Observatorio del Roque de los Muchachos, La Palma, Spain, of the Instituto de Astrof\'{\i}sica de Canarias.}

General support for this programme has been shown by the School of Chemical and Physical Sciences of the Victoria University of Wellington; as well as the \c{C}anakkale Onsekiz Mart University, Turkey, notably Prof.\ O.\ Demircan. The Royal Astronomical Society of New Zealand, particularly its Variable Stars South section (http://www.variablestarssouth.org), was also supportive. 

It is a pleasure to express our appreciation of the high-quality and ready availability, via the Mikulski Archive for Space Telescopes (MAST), of data collected by the TESS mission. Funding for the TESS mission is provided by the NASA Explorer Program. This research has made use of the SIMBAD data base, operated at CDS, Strasbourg, France, and of NASA's Astrophysics Data System Bibliographic Services. We thank the University of Queensland for the use of collaboration software. { We are grateful for the helpful comments and guidance by the anonymous referee, which led to an improved paper.}


\section{Data availability}

The TESS photometric data used in this study are publicly available from the Barbara A. Mikulski Archive for Space Telescopes (MAST) portal maintained by the Space Telescope Science Institute.  The BVR photometric data underlying this article will be shared on reasonable request to M. Blackford { and also available as supplemental material to the electronic version of the paper.}  {\sc hercules} spectroscopy can be sourced from E.\ Budding. {\sc fies} spectra can be obtained from K.\ Pavlovski, J.\ Southworth, or the NOT archive at \texttt{https://www.not.iac.es/archive/}.

\newcommand{\invisiblesection}[1]{%
  \phantomsection%
  \stepcounter{section}%
  \addcontentsline{toc}{section}{\protect\numberline{\thesection}#1}%
  }
  
\appendix
\invisiblesection{Appendix}

\begin{table}
\begin{center}
\caption{Significant peaks in VV Ori frequency spectrum.
\label{tab:App1}}
\begin{tabular}{rrr}
\hline
Freq. d$^{-1}$          & {Ampl. (mmag)}                & {Phase (rad)}  \\
\hline
$ 0.0042 \pm 0.0000$    & $11.4053 \pm  0.0047$         & $0.7062 \pm  0.0001$  \\ 
$0.1166 \pm 0.0001$     & $1.1453 \pm 0.0047$           & $0.7511 \pm 0.0007$  \\
$0.2353 \pm 0.0010$     & $0.1014 \pm 0.0047$           & $0.7168 \pm 0.0074$  \\
$0.3087 \pm 0.0012$     & $0.0816 \pm 0.0047$           & $0.6794 \pm 0.0092$  \\
$0.4459 \pm 0.0015$     & $0.0674 \pm 0.0047$           & $0.0007 \pm 0.0112$  \\
$0.5902 \pm 0.0012$     & $0.0862 \pm 0.0047$           & $0.3041 \pm 0.0087$  \\
$0.6696 \pm 0.0006$     & $0.1589 \pm 0.0047$           & $0.6249 \pm 0.0047$  \\
$0.7776 \pm 0.0013$     & $0.0797 \pm 0.0047$           & $0.4689 \pm 0.0095$  \\
$0.8196 \pm 0.0011$     & $0.0948 \pm 0.0047$           & $0.7560 \pm 0.0079$  \\
$1.1574 \pm 0.0016$     & $0.0615 \pm 0.0047$           & $0.1898 \pm 0.0122$  \\
$1.3075 \pm 0.0003$     & $0.2899 \pm 0.0047$           & $0.6053 \pm 0.0026$  \\
$1.4047 \pm 0.0007$     & $0.1513 \pm 0.0047$           & $0.4425 \pm 0.0050$  \\
$1.6139 \pm 0.0016$     & $0.0631 \pm 0.0047$           & $0.1496 \pm 0.0119$  \\
$1.9836 \pm 0.0015$     & $0.0683 \pm 0.0047$           & $0.3762 \pm 0.0110$  \\
$2.5417 \pm 0.0009$     & $0.1099 \pm 0.0047$           & $0.9549 \pm 0.0069$  \\
$2.7875 \pm 0.0011$     & $0.0897 \pm 0.0047$           & $0.4184 \pm 0.0084$  \\
$2.8376 \pm 0.0004$     & $0.2318 \pm 0.0047$           & $0.8943 \pm 0.0033$  \\
$4.0018 \pm 0.0009$     & $0.1075 \pm 0.0047$           & $0.2149 \pm 0.0070$  \\
$4.7099 \pm 0.0010$     & $0.1028 \pm 0.0047$           & $0.1232 \pm 0.0073$  \\
$5.3827 \pm 0.0006$     & $0.1798 \pm 0.0047$           & $0.0331 \pm 0.0042$  \\
$5.4875 \pm 0.0014$     & $0.0731 \pm 0.0047$           & $0.9579 \pm 0.0103$  \\
$5.6319 \pm 0.0008$     & $0.1216 \pm 0.0047$           & $0.4694 \pm 0.0062$  \\
$7.1503 \pm 0.0008$     & $0.1283 \pm 0.0047$           & $0.9558 \pm 0.0059$  \\
$7.6831 \pm 0.0008$     & $0.1243 \pm 0.0047$           & $0.9236 \pm 0.0061$  \\
$7.8308 \pm 0.0003$     & $0.3680 \pm 0.0047$           & $0.8973 \pm 0.0020$  \\
$8.0721 \pm 0.0017$     & $0.0600 \pm 0.0047$           & $0.1402 \pm 0.0126$  \\
$8.3589 \pm 0.0005$     & $0.1973 \pm 0.0047$           & $0.1094 \pm 0.0038$  \\
$8.4197 \pm 0.0014$     & $0.0718 \pm 0.0047$           & $0.7448 \pm 0.0105$  \\
$8.4772 \pm 0.0011$     & $0.0930 \pm 0.0047$           & $0.5730 \pm 0.0081$  \\
$8.5029 \pm 0.0005$     & $0.1847 \pm 0.0047$           & $0.6060 \pm 0.0041$  \\
$8.7186 \pm 0.0015$     & $0.0690 \pm 0.0047$           & $0.0109 \pm 0.0109$  \\
$9.0322 \pm 0.0002$     & $0.5902 \pm 0.0047$           & $0.2577 \pm 0.0013$  \\
$9.0958 \pm 0.0013$     & $0.0753 \pm 0.0047$           & $0.3013 \pm 0.0100$  \\
$9.1773 \pm 0.0001$     & $1.4365 \pm 0.0047$           & $0.2618 \pm 0.0005$  \\
$10.0176\pm 0.0013$     & $	0.0754\pm 0.0047$           & $0.4459 \pm  0.0100$ \\ 
$10.3193\pm  0.0009$    & $	0.1093\pm  0.0047$          & $	0.8415 \pm  0.0069$  \\
$10.3798\pm  0.0001$    & $	1.2726\pm  0.0047$          & $	0.0622 \pm  0.0006$  \\
$10.4817\pm  0.0015$    & $	0.0656\pm  0.0047$          & $	0.7674 \pm  0.0115$  \\
$11.0542\pm  0.0008$    & $	0.1322\pm  0.0047$          & $	0.1982 \pm  0.0057$  \\
$11.1685\pm  0.0004$    & $	0.2789\pm  0.0047$          & $	0.1763 \pm  0.0027$  \\
$11.2012\pm  0.0003$    & $	0.3549\pm  0.0047$          & $	0.7060 \pm  0.0021$  \\
$11.7989\pm  0.0006$    & $	0.1617\pm 0.0047$           & $	0.7024 \pm  0.0047$  \\
$11.8698\pm  0.0001$    & $	0.9092\pm  0.0047$          & $	0.4085 \pm  0.0008$  \\
$13.0721\pm  0.0002$    & $	0.5211\pm  0.0047$          & $	0.6850 \pm  0.0014$  \\
$13.2177\pm  0.0002$    & $	0.5919\pm  0.0047$          & $	0.1865 \pm  0.0013$  \\
$13.7346\pm  0.0014$    & $	0.0715\pm  0.0047$          & $	0.9988 \pm  0.0105$  \\
$13.8899\pm  0.0005$    & $	0.1992\pm  0.0047$          & $	0.3467 \pm  0.0038$  \\
$14.3084\pm  0.0013$    & $	0.0772\pm  0.0047$          & $	0.0037 \pm  0.0098$  \\
$14.4155\pm  0.0009$    & $	0.1104\pm  0.0047$          & $	0.9842 \pm  0.0068$  \\
$14.5005\pm  0.0016$    & $	0.0612\pm  0.0047$          & $	0.2587 \pm  0.0123$  \\
$14.5623\pm  0.0004$    & $	0.2594\pm  0.0047$          & $	0.0344 \pm  0.0029$  \\
$15.2349\pm  0.0012$    & $	0.0842\pm  0.0047$          & $	0.6404 \pm  0.0089$  \\
$20.9609\pm  0.0009$    & $	0.1082\pm  0.0047$          & $	0.5285 \pm  0.0078$  \\
$22.3940\pm  0.0010$    & $	0.0975\pm  0.0047$          & $	0.7536 \pm  0.0077$  \\
$23.5953\pm  0.0007$    & $	0.1417\pm  0.0047$          & $	0.0482 \pm  0.0053$  \\
$25.0870\pm  0.0006$    & $	0.1600\pm  0.0047$          & $	0.3657 \pm  0.0047$  \\
$25.7319\pm  0.0020$    & $	0.0511\pm  0.0047$          & $	0.9643 \pm  0.0148$  \\
$26.2917\pm  0.0012$    & $	0.0807\pm  0.0047$          & $	0.6437 \pm  0.0093$  \\
$27.7816\pm  0.0014$    & $	0.0697\pm  0.0047$          & $	0.9514 \pm  0.0108$  \\
 \hline \\
\end{tabular}
\end{center}
\end{table}

\begin{table}
\begin{center}
\caption{Identified spectral lines for VV Ori (p $\equiv$ primary; s $\equiv$ secondary).
\label{tab:tbl-3}}
{\footnotesize
\begin{tabular}{llll}
\hline
\multicolumn{1}{l}{Species}  & \multicolumn{1}{l}{Order no.} &
\multicolumn{1}{l}{Adopted $\lambda$} & \multicolumn{1}{l}{Comment}  \\
\hline
He I         & 85        &  6678.149  &   well-defined p;  s weak, noisy \\
H$_{\alpha}$ & 87        &  6562.817  &   s noticeable in p profile     \\
He I         & 97        &  5875.340  &  well-defined p; s weak \\
Na II        & 97        &  5895.923  &  deep, narrow structured D-lines \\
Na II        & 97        &  5889.953  &   --- \\
Si III       & 99        & 5739.620   &  p \& s visible \\
C III        & 100       &  5696.00   &  p only  \\
N II         & 100       &  5679.56   &     \\
Si II        & 101       &  5639.49   &    \\
C III        & 109       &  5217.93   &    \\
O II         & 109       &  5206.73   &    \\
He I         & 112, 113  &  5047.736  &  p strong, s visible  \\
He I         & 113       &  5015.675  &   p only   \\
N II         & 114       &  5001.30   &  \\
O II         & 115       &  4942.1    &  blend   \\
C III        & 115       &  4922.14   &    \\
H$_{\beta}$  & 117       &  4861.332  &  s visible but blended      \\
C III        & 118       &  4819.63   & p blend  \\
N II       & 121         &  4718.4    & weak   \\
He I         & 120, 121  &  4713.258  & weak p \& s \\
O II         & 121       &  4699.21   &  \\
He II        & 122       &  4685.682  &  p only \\
O II         & 122       &  4674.213  &   \\
C III        & 122       &  4663.53   &  \\
C III        & 122       &  4650.160  &    \\
C III+ OII   & 123       &  4649.86   &    strong blend   \\
N II         & 123       & 4630.567   &  \\
N II         & 123       & 4621.39    &  \\
O II         & 123       & 4610.14    &  \\
O II         & 124       &  4602.11   &  \\
O II         & 124       &   4596.174 &  \\
C III        & 124       &   4593.47  &  \\
Si III       & 125       &  4574.78   &  edge of order     \\
Si III       & 125       &  4569.67   &  blend \\
C III        & 125       & 4567.872   &  \\
Si III       & 125       &  4552.65   &  blend     \\
\hline \\
\end{tabular}}
\end{center}
\end{table}

\bibliographystyle{mnras}

\bsp	
\label{lastpage}
\end{document}